\documentclass[floatfix,twocolumn,showpacs,amssymb,amsmath,prd]{revtex4}
\usepackage{graphicx}
\usepackage{dcolumn}
\usepackage{bm}

\begin{document}
\title{CMB map derived from the WMAP data through Harmonic Internal Linear Combination}
\author{Jaiseung Kim}
\email{jkim@nbi.dk}
\affiliation{Niels Bohr Institute, Blegdamsvej 17, DK-2100 Copenhagen, Denmark}
\author{Pavel Naselsky}
\affiliation{Niels Bohr Institute, Blegdamsvej 17, DK-2100 Copenhagen, Denmark}
\author{Per Rex Christensen}
\affiliation{Niels Bohr Institute, Blegdamsvej 17, DK-2100 Copenhagen, Denmark}

\date{\today}

\begin{abstract}
We are presenting an Internal Linear Combination (ILC) CMB map, in which the foreground is reduced through harmonic variance minimization. We have derived our method by converting a general form of pixel-space approach into spherical harmonic space, maintaining full correspondence. By working in spherical harmonic space, spatial variability of linear weights is incorporated in a self-contained manner and our linear weights are continuous functions of position over the entire sky. 
The full correspondence to pixel-space approach enables straightforward physical interpretation on our approach.  
In variance minimization of a linear combination map, the existence of a cross term between residual foregrounds and CMB makes the linear combination of minimum variance differ from that of minimum foreground. 
We have developed an iterative foreground reduction method, where perturbative correction is made for the cross term.
Our CMB map derived from the WMAP data is in better agreement with the WMAP best-fit $\Lambda$CDM model than the WMAP team's Internal Linear Combination map. We find that our method's capacity to clean foreground is limited by the availability of enough spherical harmonic coefficients of good Signal-to-Noise Ratio (SNR).
\end{abstract}


\maketitle 
 
\section{Introduction}
A whole-sky map contains significant amount of foreground emission from astrophysical sources.
Hence, the ability to clean foreground contamination in CMB data is of the utmost importance for CMB observations. 
In the WMAP observation, foreground were cleaned by two different methods \citep{WMAP:TT,WMAP:foreground,WMAP:3yr_TT}.
One is using external templates of the foregrounds, but using external template maps \citep{WMAP:foreground} suffers from dubious extrapolation and template noise of higher level than the WMAP \citep{Tegmark:foreground_presence}. The other is using the Internal Linear Combination (ILC) method \citep{WMAP:3yr_TT,Eriksen:ILC}, where the linear weight for each frequency channel is chosen to minimize the variance of the linear combination of multi-frequency maps, therefore minimizing residual foreground. To take into account the spatial variability, the WMAP team defined twelve disjoint regions, where distinct linear weights are assumed for each region. In spite of many merits of the ILC method, it has important limits: First, the definition of disjoint regions requires external information and the use of disjoint regions brings about discontinuities. Second, there exists
a cross term between the residual foreground and CMB, which makes the variance minimization proceed as to maximize the cancellation between the residual foreground and CMB \citep{WMAP:3yr_TT}.
For the solution of the first problem, we have carried out the variance minimization entirely in spherical harmonic space, where the spatial variability of linear weights can be incorporated in a self-contained and seamless manner. 
For the solution of the second problem, we have developed an iterative foreground reduction method, where perturbative correction is made for the cross term. Simulations confirmed that our iterative method reconstructs the CMB with stability and reliability. 
We have also applied our method to the WMAP data and obtained a foreground-reduced CMB map. 

The outline of this paper is as follows. 
In Sec. \ref{multifrequency}, we discuss briefly the foreground reduction method with multi-frequency maps, minimum variance principle and the choice of general form for linear weights.
In Sec. \ref{spherical_space}, we derive equations in spherical harmonic space, whose solutions correspond to linear weights of minimum foreground.
We present an iterative foreground reduction method in Sec. \ref{bias} and simulation results in Sec. \ref{simulation}.
The result of application to the WMAP three year and five year data are presented in Sec. \ref{WMAP} and \ref{WMAP5YR} respectively.
We discuss computational issues in Sec. \ref{computation} and conclude this investigation in Sec. \ref{conclusion}. 
In appendix \ref{matrix_solution}, the equation for minimum foreground is put in matrix notation and the solution is presented in the form of matrix operations.
In appendix \ref{supprression}, we show that the cross term between residual foreground and CMB causes the suppression on the lowest multipole powers of an Internal Linear Combination (ILC) map.
In appendix \ref{comparison}, we make a brief comparison by summarizing the advantages and disadvantages of ILC variants and the template-fitting method.
\section{Foreground reduction with multi-frequency maps}
\label{multifrequency}
With neglect of pixel noise, a thermodynamic temperature map at a frequency  $\nu_i$ and pixel $\mathbf x$ is as follows:
\begin{eqnarray}
T(\mathbf x,\nu_i)=T_{\mathrm{cmb}}(\mathbf x)+T_{\mathrm{fg}}(\mathbf x,\nu_i),
\label{T}
\end{eqnarray}
where $T_{\mathrm{cmb}}(\mathbf x)$ and $T_{\mathrm{fg}}(\mathbf x,\nu_i)$ are CMB signal and the composite foreground signal respectively.
A natural choice for the estimator of the CMB map is a linear combination of multi-frequency maps, which is as follows: 
\[\sum_{i} {w_i(\mathbf x)}\,T(\mathbf x,\nu_i).\]
To keep the CMB unchanged, a contraint is given such that the sum of linear weights over frequency channels is equal to unity: 
\begin{eqnarray}
\sum_{i} {w_i(\mathbf x)}=1. \label{nomarlization}
\end{eqnarray}
With Eq. \ref{T} and \ref{nomarlization}, it is straightforward to show that
\begin{eqnarray}
\sum_{i} {w_i(\mathbf x)}\;T(\mathbf x,\nu_i)=T_{\mathrm{cmb}}(\mathbf x)+\sum_{i} w_i(\mathbf x)\;T_{\mathrm{fg}}(\mathbf x,\nu_i)\label{ilc_pixel}.
\end{eqnarray}

We can make the foreground signal in Eq. \ref{ilc_pixel} vanish, if the linear weights are chosen such that : 
\begin{eqnarray}
\sum_{i} w_i(\mathbf x)\;T_{\mathrm{fg}}(\mathbf x,\nu_i)=0. \label{w_constraint}
\end{eqnarray}
Since we have no information on $T_{\mathrm{fg}}(\mathbf x,\nu_i)$, we need some function to maximize or minimize, which will lead us toward such linear weights. 
One of such powerful methods is variance minimization of the linear combination map \citep{Eriksen:ILC,WMAP:3yr_TT}.
It can be shown that the variance of a linear combination map is
\begin{eqnarray}
\sigma^2&=&\left\langle \left(\sum_{i} {w_i(\mathbf x)}\;T(\mathbf x,\nu_i)\right)^2\right\rangle\label{sigma_ilc}\\
&\approx&C^2+2\left\langle T_{\mathrm{cmb}}(\mathbf x)\,\sum_{i} {w_i(\mathbf x)}\;T_{\mathrm{fg}}(\mathbf x,\nu_i)\right\rangle\nonumber\\
&&+\left\langle \left(\sum_{i} {w_i(\mathbf x)}\;T_{\mathrm{fg}}(\mathbf x,\nu_i)\right)^2\right\rangle\nonumber
\end{eqnarray}
where the constant term $C^2$ is the variance of CMB and therefore, independent of the choice of linear weight.
Though the cross term $2\left\langle T_{\mathrm{cmb}}(\mathbf x)\,\sum_{i} {w_i(\mathbf x)}\;T_{\mathrm{fg}}(\mathbf x,\nu_i)\right\rangle$ in Eq. \ref{sigma_ilc} 
vanishes, when averaged over a whole ensemble of universes, it is not necessarily zero for our single observable Universe. Hence, we assume the cross term to be small but non-zero, and will make perturbative correction for it (see Sec. \ref{bias}). For now, we neglect the cross term. 

The linear weights, which yield a foreground-free map, are functions of the frequency spectrum of foreground components. Since the frequency spectrum varies over sky (see \citep{Foreground:Spectra} for a recent treatment), the linear weights should possess spatial variability. To accommodate the spatial variability of linear weights, the WMAP team defined twelve disjoint regions in the WMAP three year ILC (WILC3YR) construction, where distinct values of linear weights are assumed for each region. The linear weights of WILC3YR have the form $w_{ij}$, where $i$ and $j$ denote a frequency channel and a region index. 
Though the WMAP team used regions of smoothed boundaries in the final map making, there still exist intrinsic discontinuities from the use of disjoint regions in variance minimization, which may even create artificial peculiarities.

To reflect the varying powers of foregrounds on different angular scales, linear weights contrived by Tegmark et al. has multipole dependency as well \citep{Tegmark:CMB_map}, and have the form $w^{ij}_{l}$.
We can easily show that optimal linear weights should possess $m$ dependency as well as $l$ dependency. For illustrative purposes, let's consider two frequency channel observation and assume the signal to consist of CMB and one foreground component only.
The spherical harmonic coefficient of $i$th channel is given by
$a^i_{lm}=a^{\mathrm{cmb}}_{lm}+a^{i,\mathrm{fg}}_{lm}$,
where $a^{i,\mathrm{fg}}_{lm}$ denotes the spherical harmonic coefficient of a foreground at $i$th channel.
Keeping the CMB signal unchanged, we assign a linear weight $w$ and $(1-w)$ to the frequency channel 1 and 2 respectively.
Then, the spherical harmonic coefficient of a linear combination map is given by
\begin{eqnarray*}
w\,a^1_{lm}+(1-w) a^2_{lm}=a^{\mathrm{cmb}}_{lm}+w a^{1,\mathrm{fg}}_{lm}+(1-w) a^{2,\mathrm{fg}}_{lm}.
\end{eqnarray*}
Obviously the linear weight $w$ yielding a foreground-free linear combination map is
\begin{eqnarray}
w=\frac{a^{2,\mathrm{fg}}_{lm}}{a^{2,\mathrm{fg}}_{lm}-a^{1,\mathrm{fg}}_{lm}}. \label{w_rank}
\end{eqnarray}
As shown in Eq. \ref{w_rank}, optimal linear weights should possess $m$ dependency as well as $l$ dependency.

The linear combination map of minimum foreground formed with multi-frequency maps 
\begin{eqnarray}
T(\theta,\phi)=\sum_i w^i(\theta,\phi)\,T(\theta,\phi,\nu_i),\label{wT}
\end{eqnarray}
can be rewritten in the spherical harmonic space, using the Clebsch-Gordon relation as:
\begin{eqnarray}
\lefteqn{a_{LM}=}\label{a_LM}\\
&&(-1)^M\sqrt{\frac{2L+1}{4\pi}}\sum_{l m} \sum_{l' m'}
\sqrt{(2l+1)(2l'+1)}\nonumber\\
&&\times \left(\begin{array}{ccc}l&l' &L\\m&m'&-M\end{array}\right)
\left(\begin{array}{ccc}l&l'&L\\0&0&0\end{array}\right)
\sum_{i} w^i_{lm}\,a^i_{l' m'}\nonumber,
\end{eqnarray}
where 
\begin{eqnarray}
a_{LM}&=&\int Y^*_{LM}(\theta,\phi)\,T(\theta,\phi)\,d\Omega,\nonumber\\
w^i_{l m}&=&\int Y^*_{l m}(\theta,\phi)\,w^i(\theta,\phi)\,d\Omega,\nonumber\\
a^i_{l' m'}&=&\int Y^*_{l' m'}(\theta,\phi)\,T(\theta,\phi,\nu_i)\,d\Omega.\nonumber
\end{eqnarray}
The constraint $\sum_i w^i(\theta,\phi)=1$ imposed to preserve the CMB signal is expressed in spherical harmonic space as follows:
\begin{eqnarray}
\sum_i w^i_{00}&=&\sqrt{4\pi},\label{w00}\\
\sum_i w^i_{lm}&=&0.\;\;\;(l> 0)\label{wlm}
\end{eqnarray}
We can see that linear weights $w^i_{lm}$ in Eq. \ref{a_LM} possess $m$ dependency as well as $l$ dependency.
Since Eq. \ref{a_LM} is equivalent to \ref{wT}, physical interpretation on Eq. \ref{a_LM}
is quite straightforward and we base our approach on Eq. \ref{a_LM}.
\section{Determination of linear weights}
\label{spherical_space}
Through variance minimization, we are going to derive equations leading toward the linear weights of minimum foreground.
Since the function $w^i(\theta,\phi)$ is real-valued, $w^i_{lm}$ obeys the reality condition $w^i_{l\,-m}=(-1)^m {w^i_{lm}}^*$. 
Therefore, only $w^i_{lm}$ ($m\ge 0$) needs to be determined.
It is computationally convenient to accommodate the reality condition by defining real-valued spherical harmonic coefficients
$\tilde{w}^i_{lm}$ as $\mathrm{Re}[w^i_{lm}]$, $\mathrm{Im}[w^i_{lm}]$ for $m\ge0$, $m<0$ respectively.
The constraints on $\tilde{w}^i_{lm}$ derived from Eq. \ref{w00} and \ref{wlm} are as follows:
\begin{eqnarray}
\sum_i \tilde{w}^i_{00}&=&\sqrt{4\pi},\label{w00_constraint}\\
\sum_i \tilde{w}^i_{lm}&=&0\;\;\;(l>0)\label{wlm_constraint}.
\end{eqnarray}

The linear weights of minimum foreground minimize the variance $\sum_{LM} |a_{LM}|^2$ under the constraints Eq. \ref{w00_constraint} and \ref{wlm_constraint}. The constrained minimization problem is solved conveniently via  Lagrange's undetermined multiplier method \citep{Arfken}.  
With the introduction of Lagrange's multiplier $\lambda_{lm}$, it can be shown that the variance is minimized
under the constraints Eq. \ref{w00_constraint} and \ref{wlm_constraint}, when
\begin{eqnarray}
\frac{\partial \sum\limits_{LM} |a_{LM}|^2}{\partial\,\tilde{w}^{i'}_{l'm'}}+\lambda_{00}\frac{\partial\left(-\sqrt{4\pi}+\partial\sum\limits_i \tilde {w}^i_{00}\right)}{\partial\,\tilde{w}^{i'}_{l'm'}}\nonumber\\
+\sum\limits_{l>0,m}\lambda_{lm}\frac{\partial\sum\limits_i \tilde {w}^i_{lm}}{\partial\,\tilde{w}^{i'}_{l'm'}}&=&0.\label{variance_minimized}
\end{eqnarray}
By using Eq. \ref{a_LM}, it can be shown that Eq. \ref{variance_minimized} has the following form:
\begin{eqnarray}
\sum_{ilm} \left[\alpha^{i'i}_{l'm'lm}\,\tilde{w}^i_{lm}\right]+\lambda_{l'm'}=0,\label{minimum_wlm}
\end{eqnarray}
where $\alpha^{i'i}_{l'm'lm}$ is
\begin{eqnarray}
\alpha^{i'i}_{l'm'lm}=2\mathrm{Re}\left[\sum_{LM}\tilde{\gamma}^*_{i'}(l',m',L,M)\,\tilde{\gamma}_{i}(l,m,L,M)\right],\label{alpha}
\end{eqnarray}
and  
$\tilde{\gamma}_i(l_1,m_1,l_3,m_3)$ is
\begin{eqnarray}
\left\{\begin{array}{r}
\gamma_i(l_1,m_1,l_3,m_3)+(-1)^{m_1}\gamma_i(l_1,-m_1,l_3,m_3)\\
\gamma_i(l_1,m_1,l_3,m_3)\\
\imath\left[\gamma_i(l_1,-m_1,l_3,m_3)-(-1)^{m_1}\gamma_i(l_1,m_1,l_3,m_3)\right]\end{array}\right.&&\nonumber
\end{eqnarray}
 for $m_1>0$, $m_1=0$ and $m_1<0$ respectively, and
\begin{eqnarray}
\lefteqn{\gamma_{i}(l_1,m_1,l_3,m_3)=}\label{gamma}\\
&&\sum_{l_2 m_2}(-1)^{m_3}\sqrt{\frac{(2l_1+1)(2l_2+1)(2l_3+1)}{4\pi}}\nonumber\\
&&\times \left(\begin{array}{ccc}l_1&l_2&l_3\\m_1&m_2&-m_3\end{array}\right)
\left(\begin{array}{ccc}l_1&l_2&l_3\\0&0&0\end{array}\right)
a^{i}_{l_2 m_2}.\nonumber
\end{eqnarray}
Therefore, the values of linear weights of minimum foreground can be found in terms of Lagrange's multiplier $\lambda_{l'm'}$ by solving the system of simultaneous linear equations given by Eq. \ref{minimum_wlm}.
The values of Lagrange's multiplier $\lambda_{l'm'}$ can be easily determined by making the solutions of Eq. \ref{minimum_wlm} satisfy the constraints Eq. \ref{w00_constraint} and \ref{wlm_constraint}. 
We can write Eq. \ref{w00_constraint}, \ref{wlm_constraint} and \ref{minimum_wlm} in matrix form and obtain the solution conveniently via matrix operations (For details on the solution in matrix notation, refer to Eq. \ref{w_solution}.).

\section{Perturbative correction for the cross term}
\label{bias}
There exists a non-zero correlation between foregrounds and true CMB, so called `Cosmic Covariance' \citep{Cosmic_Covariance}, which leads to a non-negligible cross term in Eq. \ref{sigma_ilc}. 
The existence of this non-negligible cross term makes the linear combination of minimum variance differ from that of minimum residual foregrounds \citep{WMAP:3yr_TT,Cosmic_Covariance}.
By noting that the cross term disappears in the absence of CMB signal, we have developed a perturbative method, where 
the cross term is reduced through iterations.  
However, as it was shown in \citep{Cosmic_Covariance}, this approach is only effective down to the level of `Cosmic Covariance' \citep{Cosmic_Covariance}.

Consider the quantity $T(\mathbf x,\nu_i)-\tilde T^{j-1}_{\mathrm{cmb}}(\mathbf x)$, where $\tilde T^{j-1}_{\mathrm{cmb}}(\mathbf x)$ is our best guess CMB map from the $(j-1)$th iteration with 
\begin{eqnarray}
\tilde T^{0}_{\mathrm{cmb}}(\mathbf x)=0 \label{guess_T0}.
\end{eqnarray}

The merit of this quantity is that CMB signal is reduced through iterations, leading to reduction of the cross term. 
We obtain linear weights $w^j_i(\mathbf x)$ of the $j$th iteration through variance minimization of 
\[\sum_i w^j_i(\mathbf x)\,\left(T(\mathbf x,\nu_i)-\tilde T^{j-1}_{\mathrm{cmb}}(\mathbf x)\right),\]
and update our best guess CMB map as follows:
\begin{eqnarray*}
\tilde T^{j}_{\mathrm{cmb}}(\mathbf x)&=&\tilde T^{j-1}_{\mathrm{cmb}}(\mathbf x)+\sum_i w^j_i(\mathbf x)\,
\left(T(\mathbf x,\nu_i)-\tilde T^{j-1}_{\mathrm{cmb}}(\mathbf x)\right).
\end{eqnarray*}
Using Eq. \ref{T} and \ref{nomarlization}, we may show that our updated CMB map is   
\begin{eqnarray}
\tilde T^{j}_{\mathrm{cmb}}(\mathbf x)=T_{\mathrm{cmb}}(\mathbf x)+\sum_i w^j_i(\mathbf x)\,T_{\mathrm{fg}}(\mathbf x,\nu_i),\label{iteration_j}
\end{eqnarray}
where  $T_{\mathrm{cmb}}(\mathbf x)$ is a true CMB map. 
Using Eq. \ref{T} and \ref{nomarlization}, we may also show that the variance of $\sum_i w^j_i(\mathbf x)\,\left(T(\mathbf x,\nu_i)-\tilde T^{j-1}_{\mathrm{cmb}}(\mathbf x)\right)$ is
\begin{eqnarray}
\lefteqn{\langle\left(\sum_i w^j_i(\mathbf x)\,\left(T(\mathbf x,\nu_i)-\tilde T^{j-1}_{\mathrm{cmb}}(\mathbf x)\right)\right)^2\rangle}\label{variance_iteration}\\
&=&\langle\left(T_{\mathrm{cmb}}(\mathbf x)-\tilde T^{j-1}_{\mathrm{cmb}}(\mathbf x)+\sum_i w^j_i(\mathbf x)\,T_{\mathrm{fg}}(\mathbf x,\nu_i)\right)^2\rangle\nonumber\\
&=&C^2+\langle\left(\sum_i w^j_i(\mathbf x)\,T_{\mathrm{fg}}(\mathbf x,\nu_i)\right)^2\rangle\nonumber\\
&&+2\langle \left(T_{\mathrm{cmb}}(\mathbf x)-\tilde T^{j-1}_{\mathrm{cmb}}(\mathbf x)\right)\,\left(\sum_i w^j_i(\mathbf x)\,T_{\mathrm{fg}}(\mathbf x,\nu_i)\right)\rangle\nonumber,
\end{eqnarray}
where $C^2$ is a term independent of linear weights $w^j_i(\mathbf x)$.
Using Eq. \ref{iteration_j}, the cross term, which is in the last line of Eq. \ref{variance_iteration}, may be shown to be 
\[-2\langle \left(\sum_i w^{j-1}_i(\mathbf x)\,T_{\mathrm{fg}}(\mathbf x,\nu_i)\right)\,\left(\sum_i w^j_i(\mathbf x)\,T_{\mathrm{fg}}(\mathbf x,\nu_i)\right)\rangle.\]
Therefore, the cross term is getting reduced through iterations, provided that 
\[\left(\sum_i w^{j+1}_i(\mathbf x)\,T_{\mathrm{fg}}(\mathbf x,\nu_i)\right)<\left(\sum_i w^{j-1}_i(\mathbf x)\,T_{\mathrm{fg}}(\mathbf x,\nu_i)\right).\]
In practice, the cross term converges to some non-zero value, which may be attributed to two causes. First, foreground are reduced with some error, which arises from the imperfection of the method applied (e.g. a finite number of assumed $w^i_{lm}$, imperfection of twelve disjoint regions of WILC3YR). The residual foreground related to the error are reduced barely through iterations.
Second, residual foregrounds of the $j$th iteration possesses some level of correlation with that of the $j-1$th iteration.

\section{Application to Simulated Data}
\label{simulation}
We have generated simulated data as follows:
\begin{eqnarray}
a^{i}_{lm}=\frac{a^{i,\Lambda}_{lm}}{B^{i,\Lambda}_{l}}-\frac{a^{\mathrm{ILC}}_{lm}}{B^{\mathrm{ILC}}_{l}}+a^{\mathrm{sim}}_{lm} \label{alm_simulation}.
\end{eqnarray}
$a^{i,\Lambda}_{lm}$ is the spherical harmonic coefficients of the WMAP band maps at $i$th channel from the LAMBDA site, and  $a^{\mathrm{ILC}}_{lm}$ and $a^{\mathrm{sim}}_{lm}$ are those of WILC3YR and a simulated CMB map.
$B^{i,\Lambda}_{l}$ is the beam transfer functions of the WMAP $i$th channel \citep{WMAP:3yr_TT} and
$B^{\mathrm{ILC}}_{l}$ is the beam transfer function of a $1^\circ$ FWHM Gaussian beam, which is the smoothing kernel used in WILC3YR.
Our procedure for the generation of simulated data is overly conservative, because the presence of instrument noise and residual foreground in the WILC3YR makes some foreground and instrument noise double-counted.
Using Eq. \ref{alm_simulation}, we have generated four hundred simulated data set and carried out foreground reduction on them.

\begin{figure}[htb!]
\includegraphics[scale=.5]{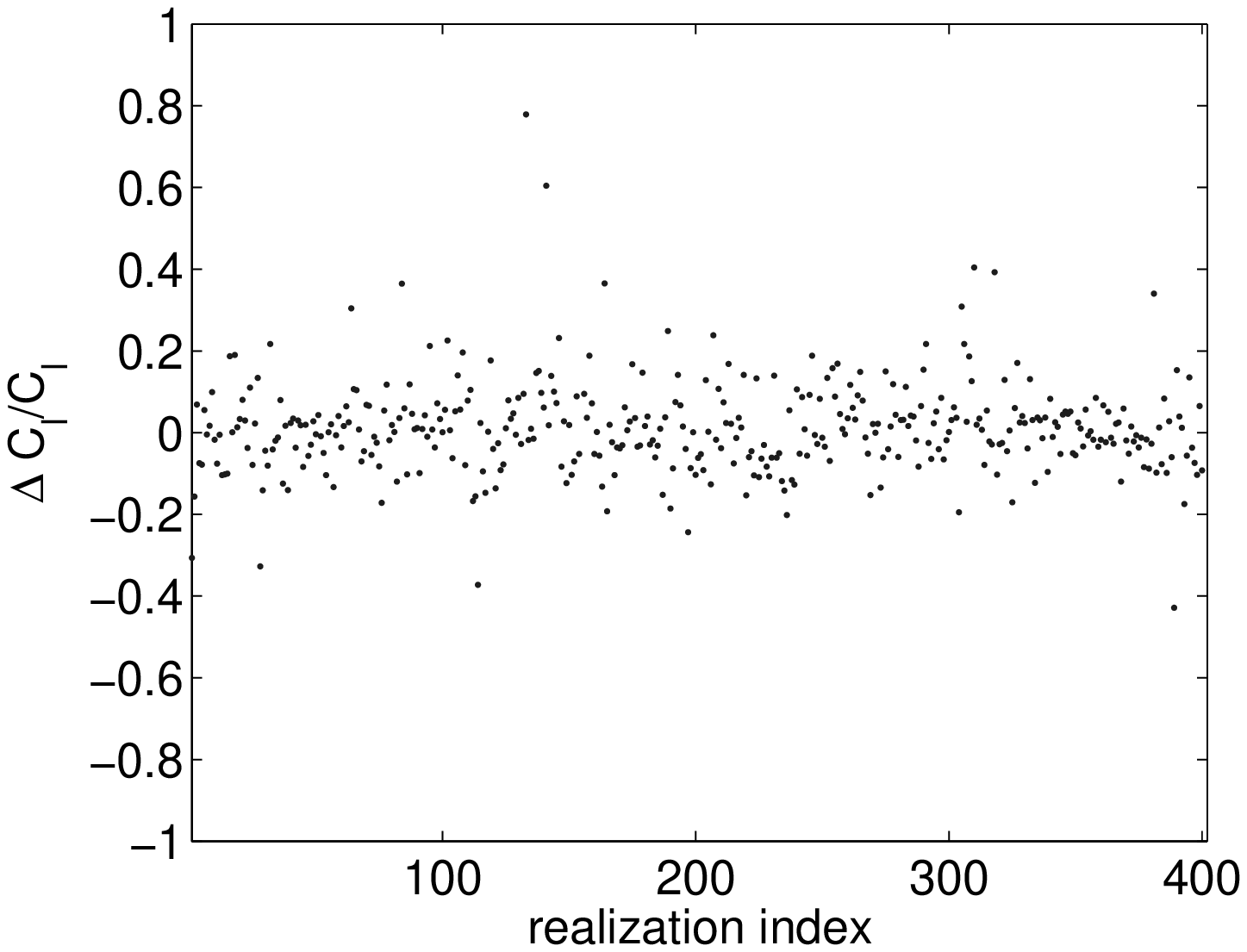}
\caption{$l=2$}
\label{Cl2}
\end{figure}

\begin{figure}[htb!]
\includegraphics[scale=.5]{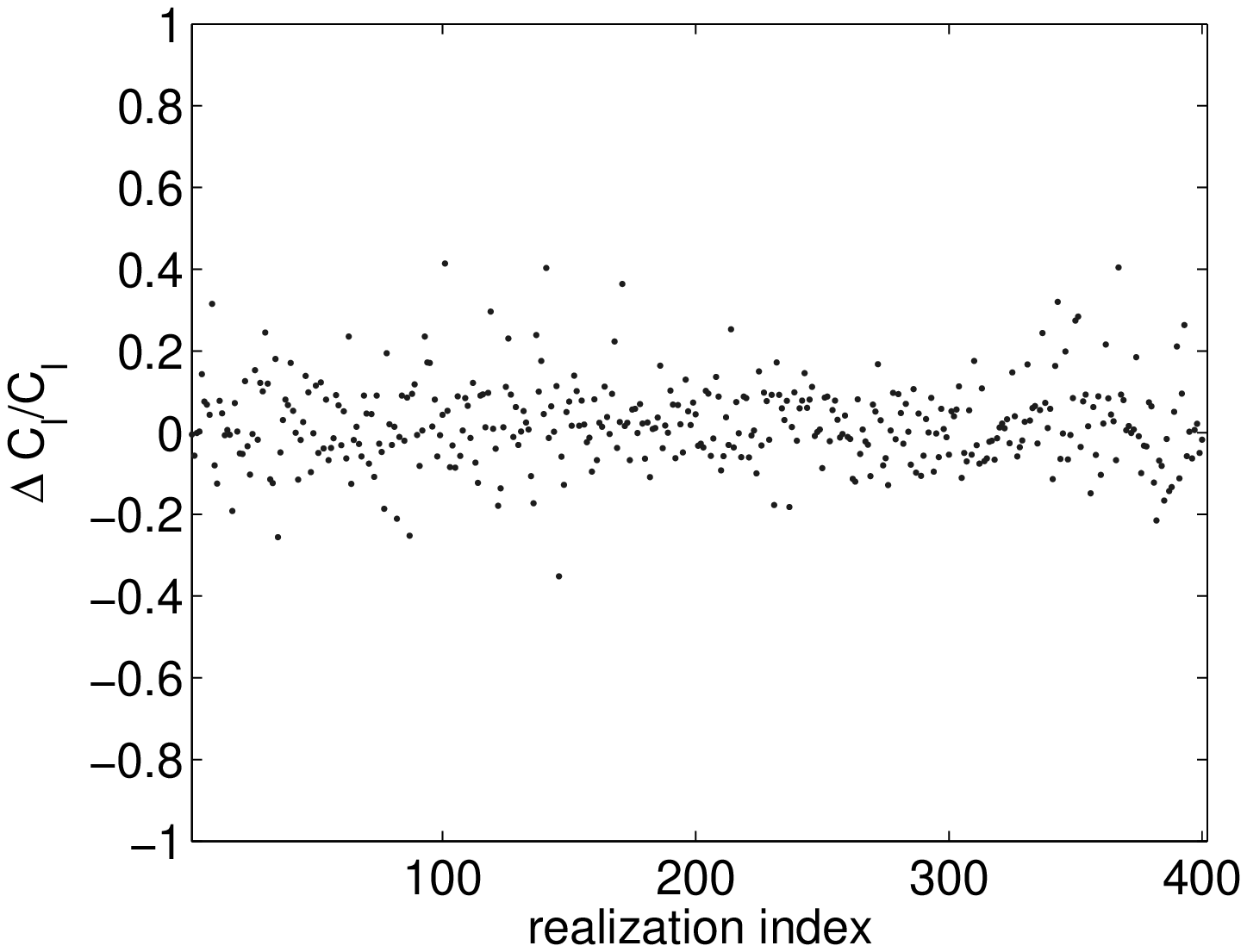}
\caption{$l=3$}
\label{Cl3}
\end{figure}

\begin{figure}[htb!]
\includegraphics[scale=.5]{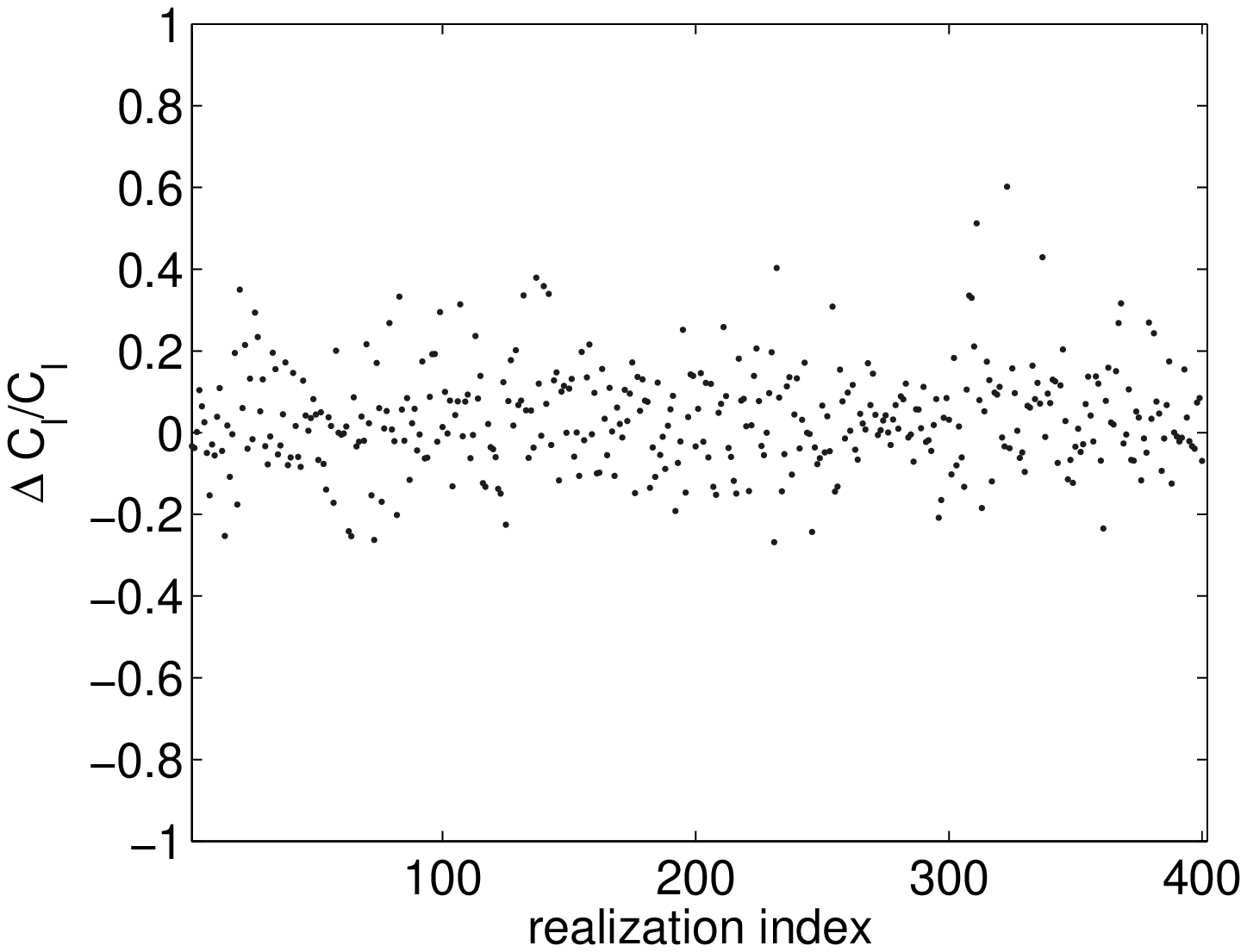}
\caption{$l=4$}
\label{Cl4}
\end{figure}

\begin{figure}[htb!]
\includegraphics[scale=.5]{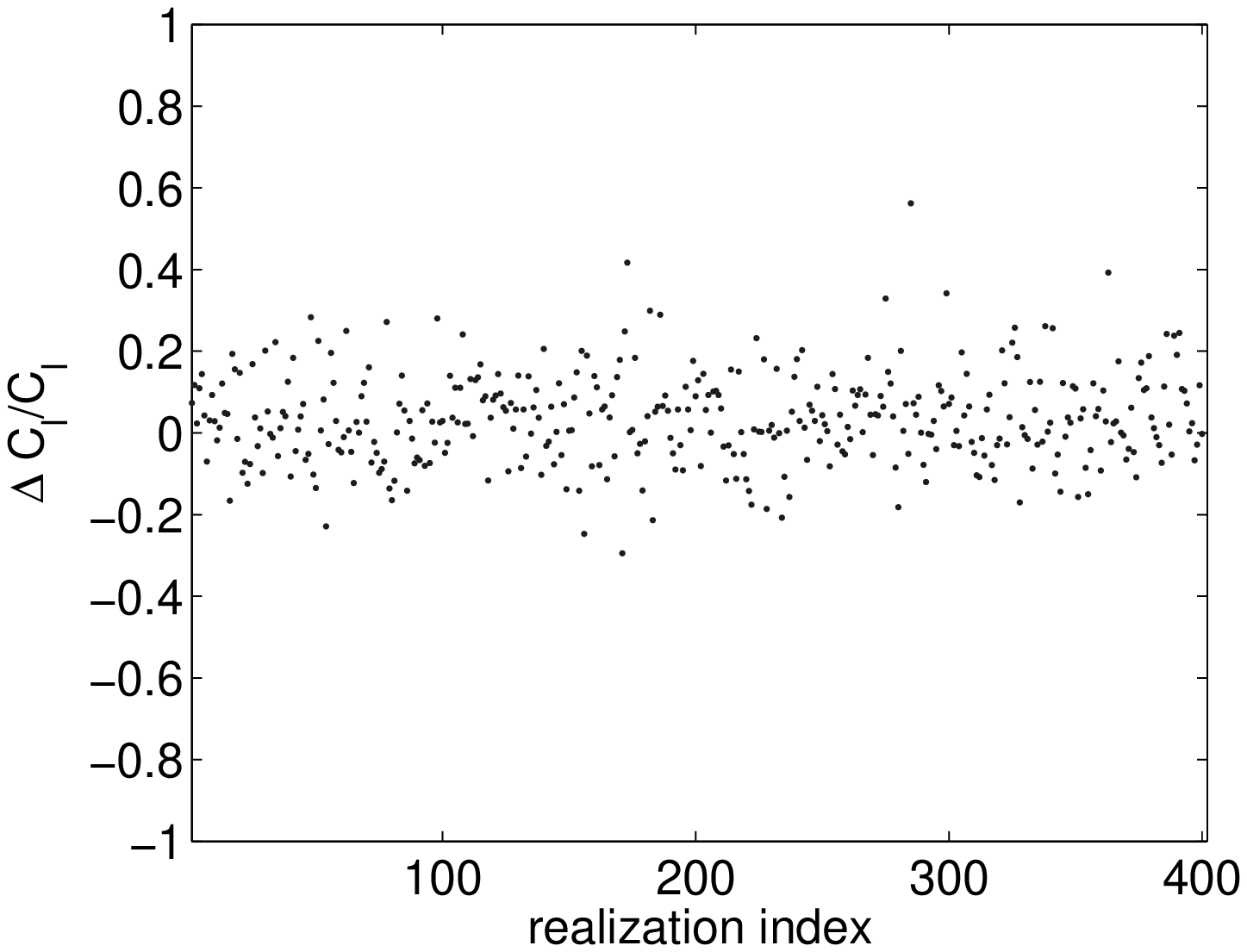}
\caption{$l=5$}
\label{Cl5}
\end{figure}

\begin{figure}[htb!]
\includegraphics[scale=.5]{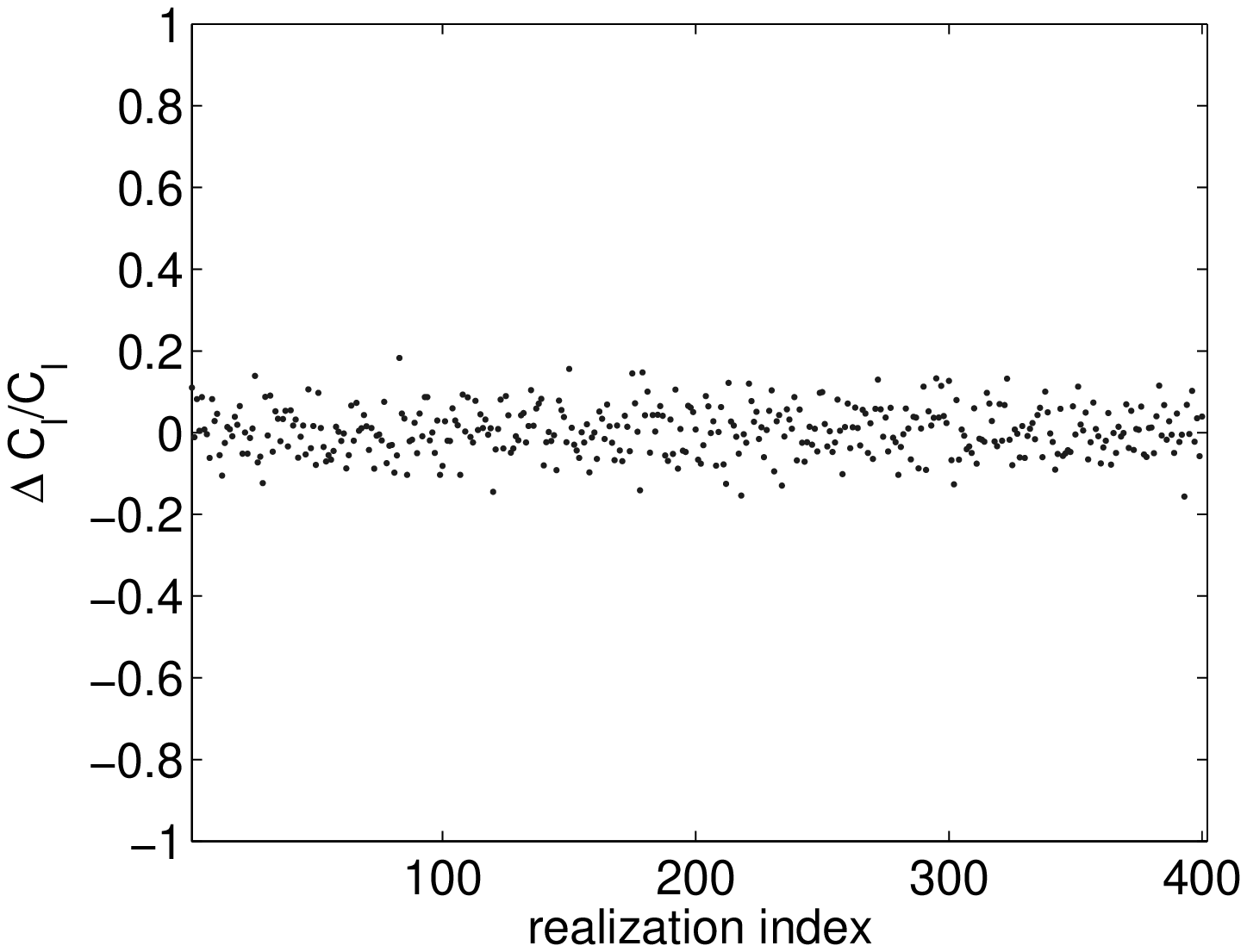}
\caption{$l=20$}
\label{Cl20}
\end{figure}

When linear weights are obtained through variance minimization on noisy data, linear weights are chosen as to minimize noise 
rather than foreground. Since our simulated data are quite noisy on multipoles higher than $300$,
we have used only $a^i_{lm}\;\;(l\le 300)$ in variance minimization (i.e. summation over $l_2$ was done up to $300$ in Eq. \ref{gamma}).

We have made the assumption that linear weights of minimum foreground will be spatially coherent on small angular scales, and determined only $w_{lm}$ in a finite multipole range $0\le l\le l_{\mathrm{cutoff}}$. 
Though ideally the total number of $w^i_{lm}$ may be as high as the total number of available $a^i_{lm}$ (i.e. the number of unknowns  may be as many as the number of constraints), we found that the number of $w^i_{lm}$, which keeps the matrices in Eq. \ref{w_solution} numerically non-singular, is much smaller than the ideal case (i.e. $l_{\mathrm{cutoff}}\ll 300$) (See Section \ref{computation} for the discussion on the possible sources of numerical singularity.). 
We have increased $l_{\mathrm{cutoff}}$ until the numerical instability emerges and found that $l_{\mathrm{cutoff}}=7$ is optimal for the WMAP data.  If more $a^i_{lm}$ of good SNR were available, it would be numerically stable with higher $l_{\mathrm{cutoff}}$. 

\begin{figure}[htb!]
\includegraphics[scale=.5]{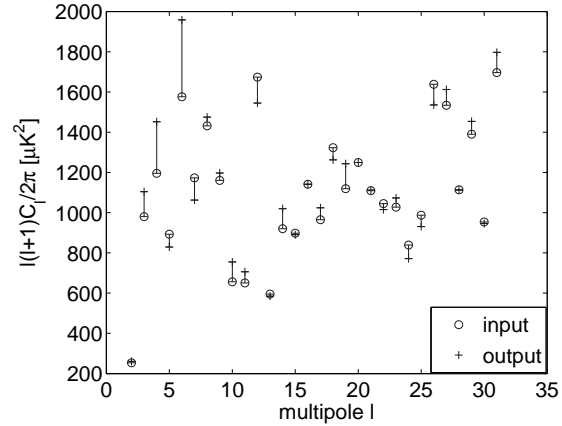}
\caption{realization \#22}
\label{sim_22}
\end{figure}

We have implemented iterative foreground reduction, which is discussed in Sec. \ref{bias}.
The cross term in the first iteration is quite significant, since the best-guess CMB map of zeroth iteration is set to zero (see Eq. \ref{guess_T0}). 
Hence, in the implementation of iterative foreground reduction, we have excluded lowest multipoles ($0\le l\le10$) in variance minimization of the first iteration for regularization purpose (see appendix \ref{supprression} for details on why exclusion of lowest multipoles reduces the effect of the cross term.). Such regularization is not necessary in succeeding iterations ($j\ge 2$).
Since it turned out that the improvement after $j=2$ iterations were negligible, we carried out only $j=2$ iterations for each simulated data set.

The power spectra discrepancy between output CMB and input CMB are shown for four hundred CMB realizations in Fig. \ref{Cl2}, \ref{Cl3}, \ref{Cl4}, \ref{Cl5} and \ref{Cl20} on multipoles ($2\le l\le 5,\;l=20$). The vertical axis denotes $(C^{\mathrm{out}}_l-C^{\mathrm{in}}_l)/C^{\mathrm{in}}_l$, where $C_l=(2l+1)^{-1}\sum_m |a_{lm}|^2$.
The horizontal axis denotes the enumerating index of four hundred CMB realizations.
In Fig. \ref{sim_22}, we show power spectra ($2\le l\le 30$) of input CMB and output CMB for the realization \#22, since the realization \#22 among four hundred realizations has the octupole and quadrupole power closest to those of the WMAP best-fit 
$\Lambda$CDM.

\section{Application to the WMAP three year data}
\label{WMAP}
\begin{figure}[htb!]
\centering\includegraphics[scale=1]{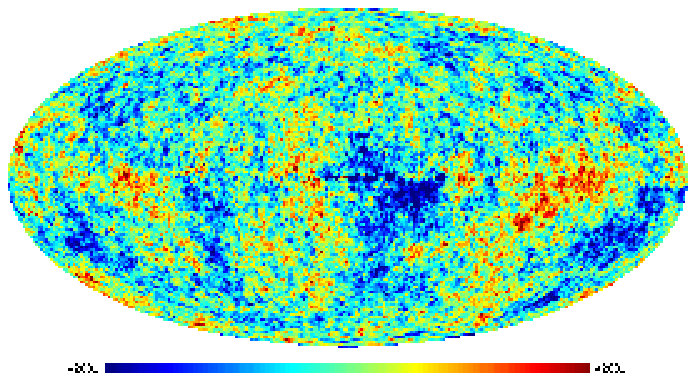}
\centering\includegraphics[scale=1]{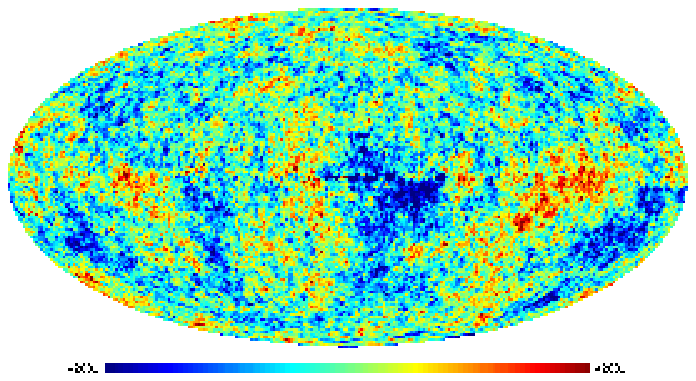}
\centering\includegraphics[scale=1]{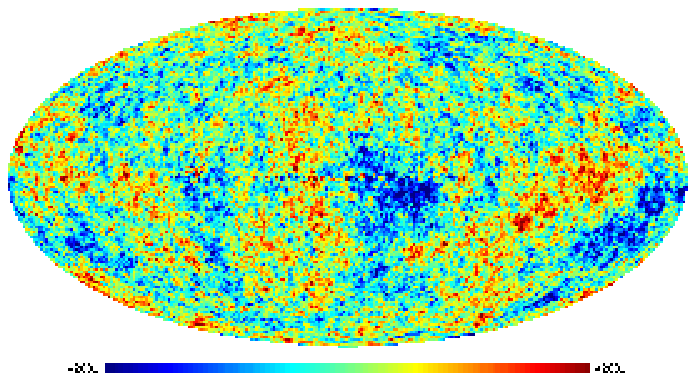}
\centering\includegraphics[scale=1]{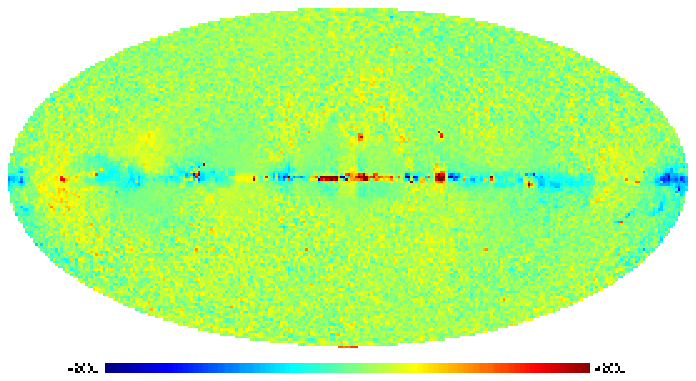}
\centering\includegraphics[scale=1]{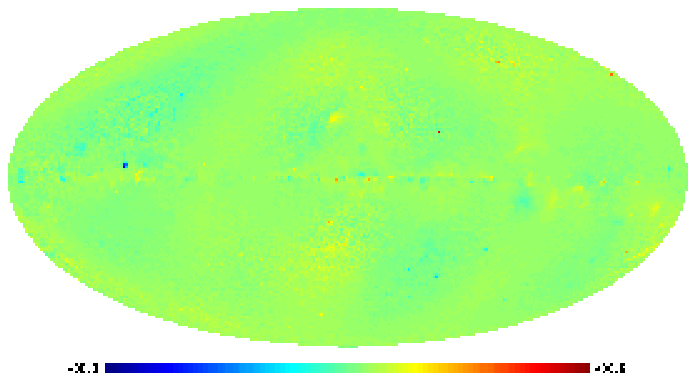}
\caption{the $1^\circ$ FWHM smoothed maps [$\mu\mathrm K$]: HILC3YR  of the zeroth iteration (top), HILC3YR  of the first iteration (second), WILC3YR (third), $\mathrm{WILC3YR}-\mathrm{HILC3YR }$ (fourth), and the difference between the zeroth and the first iteration  HILC3YR (bottom)}
\label{ilc}
\end{figure}
We have applied our foreground reduction method to the WMAP three year data from the LAMBDA site \citep{WMAP:3yr_TT}.
Spherical harmonic coefficients of the band maps have been obtained as follows: \[a^{i}_{lm}=a^{i,\Lambda}_{lm}/B^{i,\Lambda}_{l},\]
where $a^{i,\Lambda}_{lm}$ are the spherical harmonic coefficients of the WMAP three year band maps and $B^{i,\Lambda}_{l}$ are the beam transfer functions of the WMAP $i$th channel \citep{WMAP:3yr_TT}.
Just as the application to the simulated data in the previous section, the cutoff multipole $l_{\mathrm{cutoff}}$ for linear weights is set to seven, and we have used $a^i_{lm}$ in the multipole range $l\le 300$ in variance minimization (i.e. summation over $l_2$ was done up to $300$ in Eq. \ref{gamma}). 
In Fig. \ref{ilc}, our CMB map, which we call `Harmonic Internal Linear Combination map' (hereafter, HILC3YR ), is shown with the WMAP three year ILC map (WILC3YR) and the difference map. The maps in Fig. \ref{ilc} are images smoothed with $1^\circ$ FWHM beam.

\begin{figure}[htb!]
\includegraphics[scale=.55]{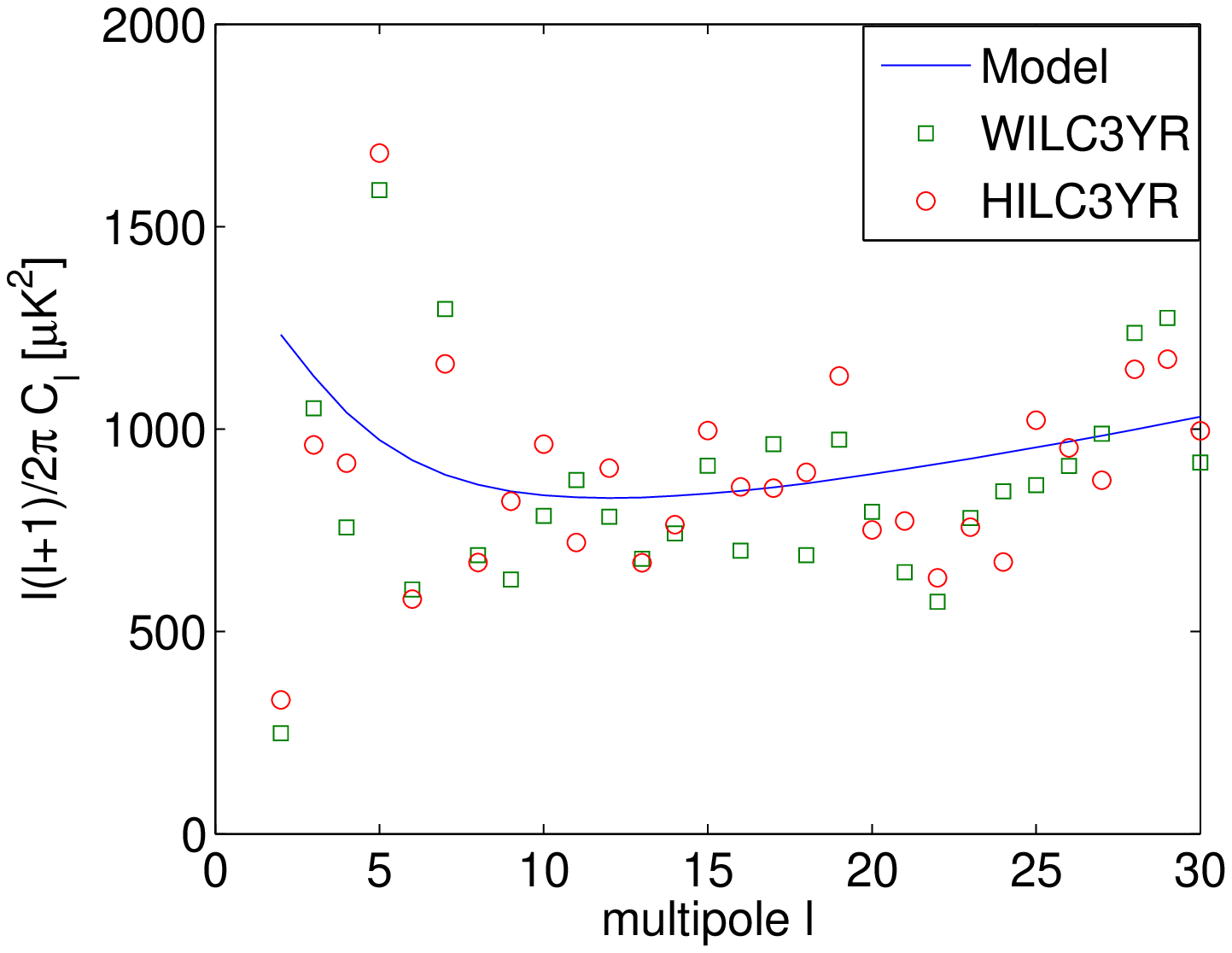}
\includegraphics[scale=.55]{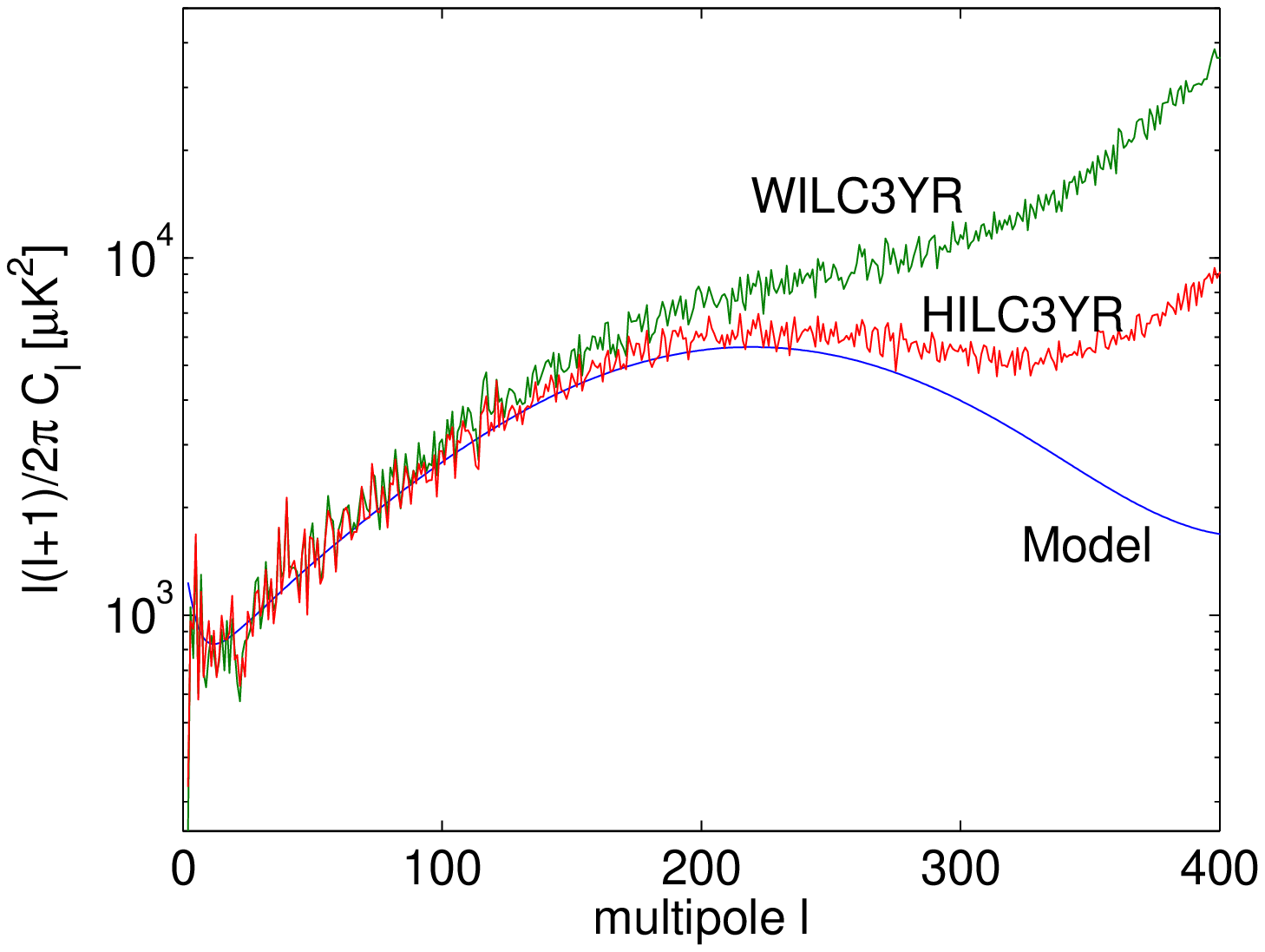}
\caption{Power spectra of HILC3YR, WILC3YR and the WMAP 3 year best-fit $\Lambda$CDM model}
\label{Cl_wmap}
\end{figure}

Power spectra estimate on HILC3YR and WILC3YR are made by computing $C_l=(2l+1)^{-1}\sum_m |a_{lm}|^2$, which are shown 
with the WMAP best-fit $\Lambda$CDM model in Fig. \ref{Cl_wmap}.
As shown in Fig. \ref{Cl_wmap}, HILC3YR makes better agreement with the WMAP best-fit $\Lambda$CDM model than WILC3YR (e.g. The first Doppler acoustic peak is visible in the HILC3YR power spectrum around $l\sim220$).
The huge excess power of WILC3YR and HILC3YR on high multipoles ($l> 300$) is attributed to pixel noise.
In the multipole range ($200< l< 300$), where point sources are dominant over other sources \citep{Tegmark:Foreground}, HILC3YR makes relatively good agreement with the model, while there is significant discrepancy between WILC3YR and the model. This may indicates relatively effective reduction of point sources in HILC3YR.
The absolute value of the power spectra difference between WILC3YR and HILC3YR is also shown in Fig. \ref{Cl_wmap}.
Through polynomial-fitting in the multipole range ($l>100$), we have investigated the multipole dependency of the 
the power spectra difference between HILC3YR and WILC3YR. Considering the multipole dependency, we may, with some caution, attribute the power difference to residual point sources and pixel noise.
\begin{figure}[htb!]
\centering\includegraphics[scale=1]{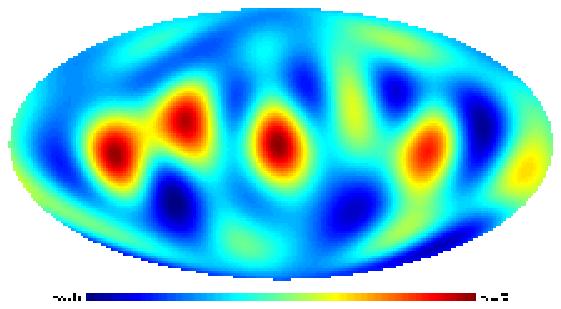}
\centering\includegraphics[scale=1]{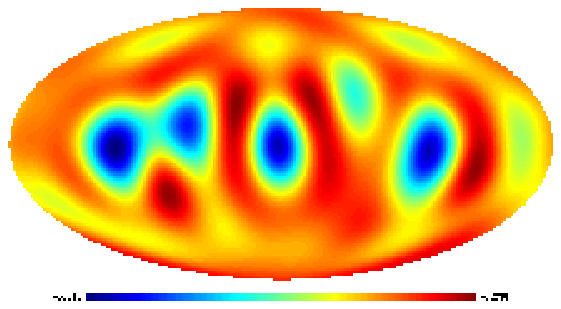}
\centering\includegraphics[scale=1]{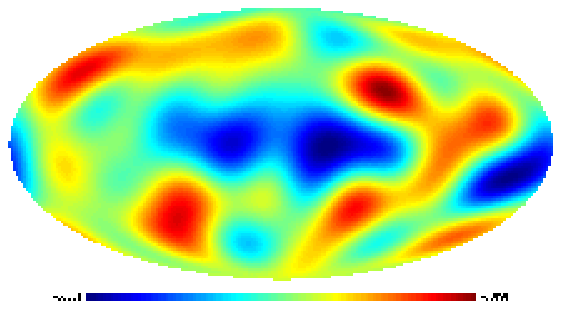}
\centering\includegraphics[scale=1]{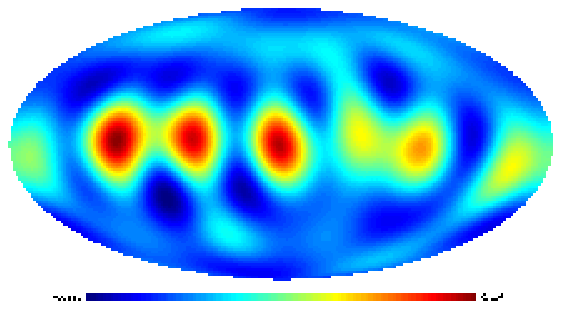}
\centering\includegraphics[scale=1]{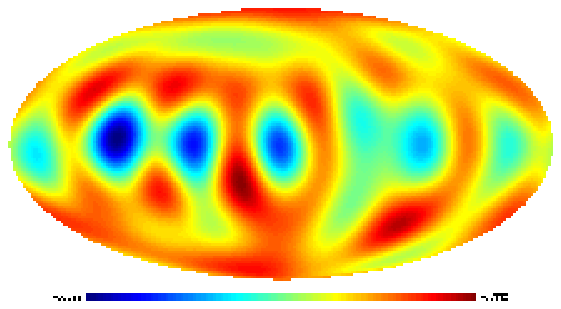}
\caption{The HILC3YR linear weight for the K, Ka, Q, V, and W band map (from top to bottom)}
\label{W_HILC}
\end{figure}
We have also investigated the linear weights of HILC3YR, and found that HILC3YR gets contribution more from the V band map than the W band map. Hence, the less noise level of HILC3YR does not mean that HILC3YR prefers blindly the least noise channel at the sacrifice of foreground reduction.
The linear weights of HILC3YR, which are continuous over entire sky, are shown in Fig. \ref{W_HILC}.
\begin{figure}[htb!]
\centering\includegraphics[scale=.4]{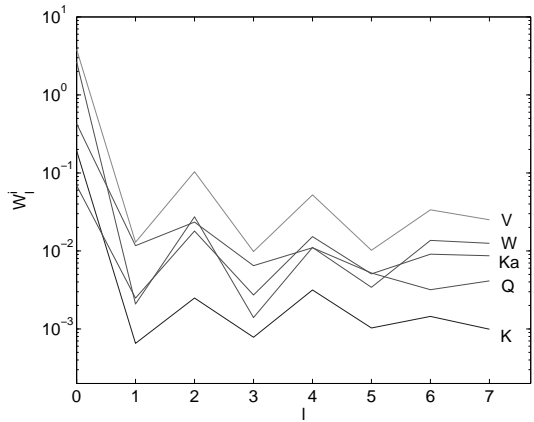}
\caption{Variance of the HILC3YR linear weights}
\label{Wl}
\end{figure}
We have computed the variance of our linear weights by $W^i_l=(2l+1)^{-1}\sum_m |w^i_{lm}|^2$ 
to quantify the spatial variation of our linear weights on different angular scales.
As shown in Fig. \ref{Wl}, $W^i_{l}$ tends to decrease with increasing multipole, with $W^i_{0}$ being the highest. It is not difficult to see from the tail pattern that there will be some non-zero $w^i_{lm}$ on multipoles higher than our assumed cutoff multipole $l=7$. 
These unaccounted $w^i_{lm}$ ($l>7$) may be partially responsible for the residual foregrounds, which is visible around the galactic plane in Fig. \ref{ilc}. 
However, it is unlikely that low multipole anisotropies are affected significantly by the residual foreground around the Galactic plane.

\begin{table}[htb!]
\centering
\caption{quadrupole and octupole powers}
\begin{tabular}{cccc}
\hline\hline 
Measurement &  $\delta T^2_2$[$\mu\mathrm{K}^2$] & p-value &$\delta T^2_3$[$\mu\mathrm{K}^2$]\\ 
\hline
WMAP best-fit $\Lambda$CDM model &  1250& $\ldots$ & 1143\\ 
Hinshaw et al. cut sky &  211.0 & 2.6\%& 1041\\
WMAP team's ILC (WILC3YR) & 248.6& 3.7\%&1051.5\\
Tegmark et al. (TCM3YR) & 209.6 & 2.5\%&1037.8\\ 
HILC3YR &  331 & 6.8\% &961\\ 
\hline 
\end{tabular}
\label{low_multipole_power}
\end{table}
In Table \ref{low_multipole_power}, the quadrupole and octupole power of HILC3YR are shown with those of the WMAP best-fit $\Lambda$CDM model and other measurements.
The p-value in Table \ref{low_multipole_power} denotes the chance of having the quadrupole power lower than the measurements on the left. As shown in Table \ref{low_multipole_power}, the p-value of HILC3YR quadrupole is almost twice that of WILC3YR.
The preferred axis of arbitrary multipoles can be quantified by finding the axis $\hat{\mathbf n}$, which maximizes the angular momentum dispersion of the corresponding multipole \citep{Tegmark:Alignment}.
It was noticed that the preferred axis of quadrupole anisotropy is close to being in alignment with that of octupole \citep{Tegmark:Alignment}. The angular separation between the preferred axis of quadrupole and octupole of HILC3YR is shown with that of other foreground-reduced maps in Table \ref{alignment}. The p-values denotes the probability of the angular separation lower than the measurements, provided that the direction of a preferred axis is random.
\begin{table}[htb!]
\centering
\caption{quadrupole-octupole alignment}
\begin{tabular}{ccc}
\hline\hline 
Maps &  $\theta_{23}$ & p-value\\ 
\hline
WMAP team's ILC (WILC3YR) & $5^\circ.9$&0.53\%\\
Tegmark et al. (TCM3YR) &  $13^\circ.2$& 2.65\%\\ 
HILC3YR &  $13^\circ.1$&2.60\%\\ 
\hline 
\end{tabular}
\label{alignment}
\end{table}

\section{Application to the WMAP five year data}
\label{WMAP5YR}
\begin{figure}[htb!]
\centering\includegraphics[scale=1]{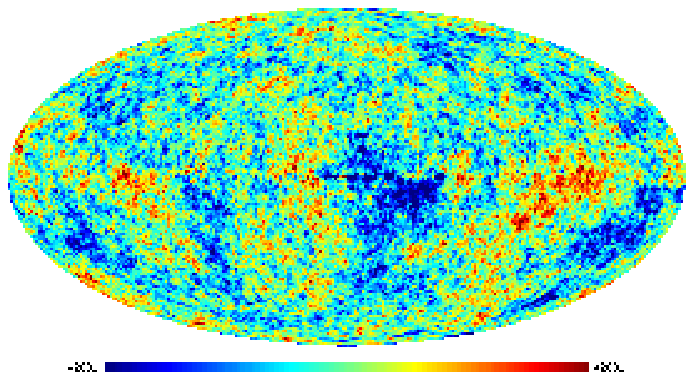}
\centering\includegraphics[scale=1]{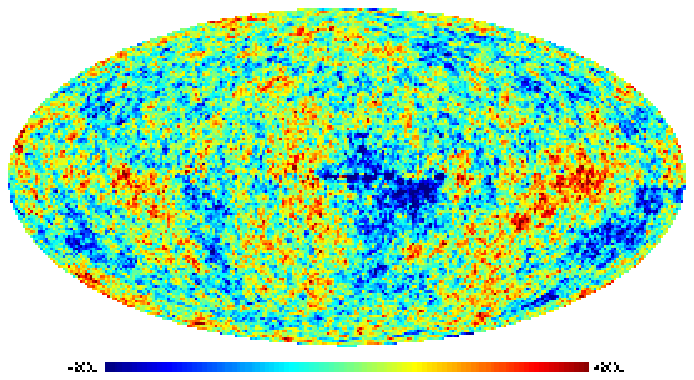}
\centering\includegraphics[scale=1]{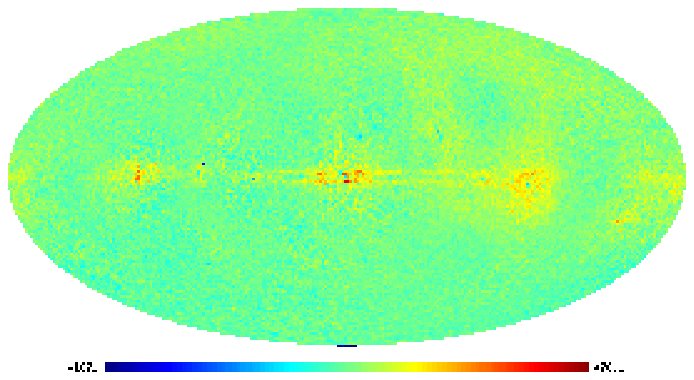}
\caption{HILC3YR  (top), HILC5YRa  (middle), HILC3YR - HILC5YRa (bottom)}
\label{HILC5YR1}
\end{figure}
\begin{figure}[htb!]
\centering\includegraphics[scale=.25]{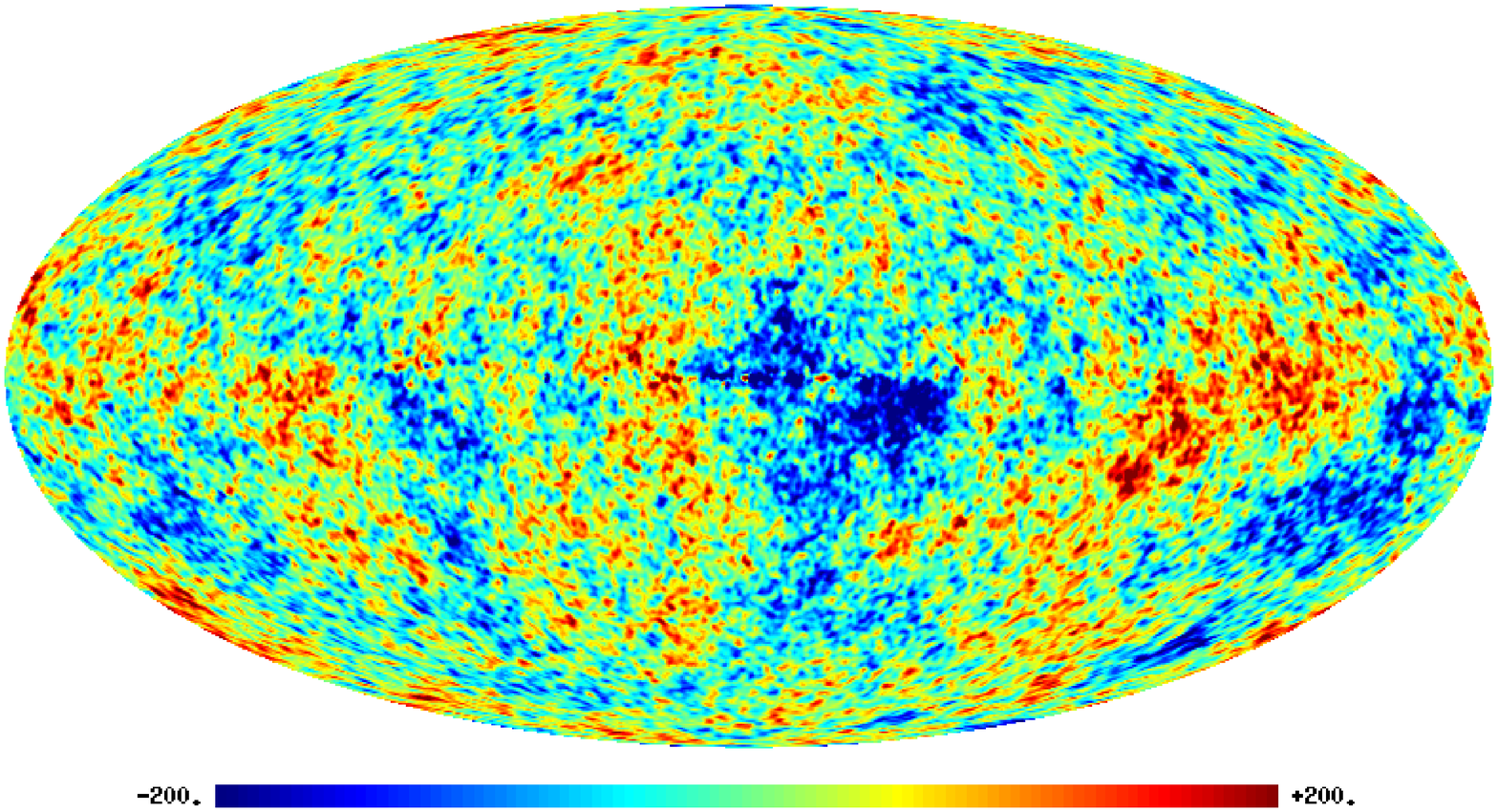}
\centering\includegraphics[scale=.25]{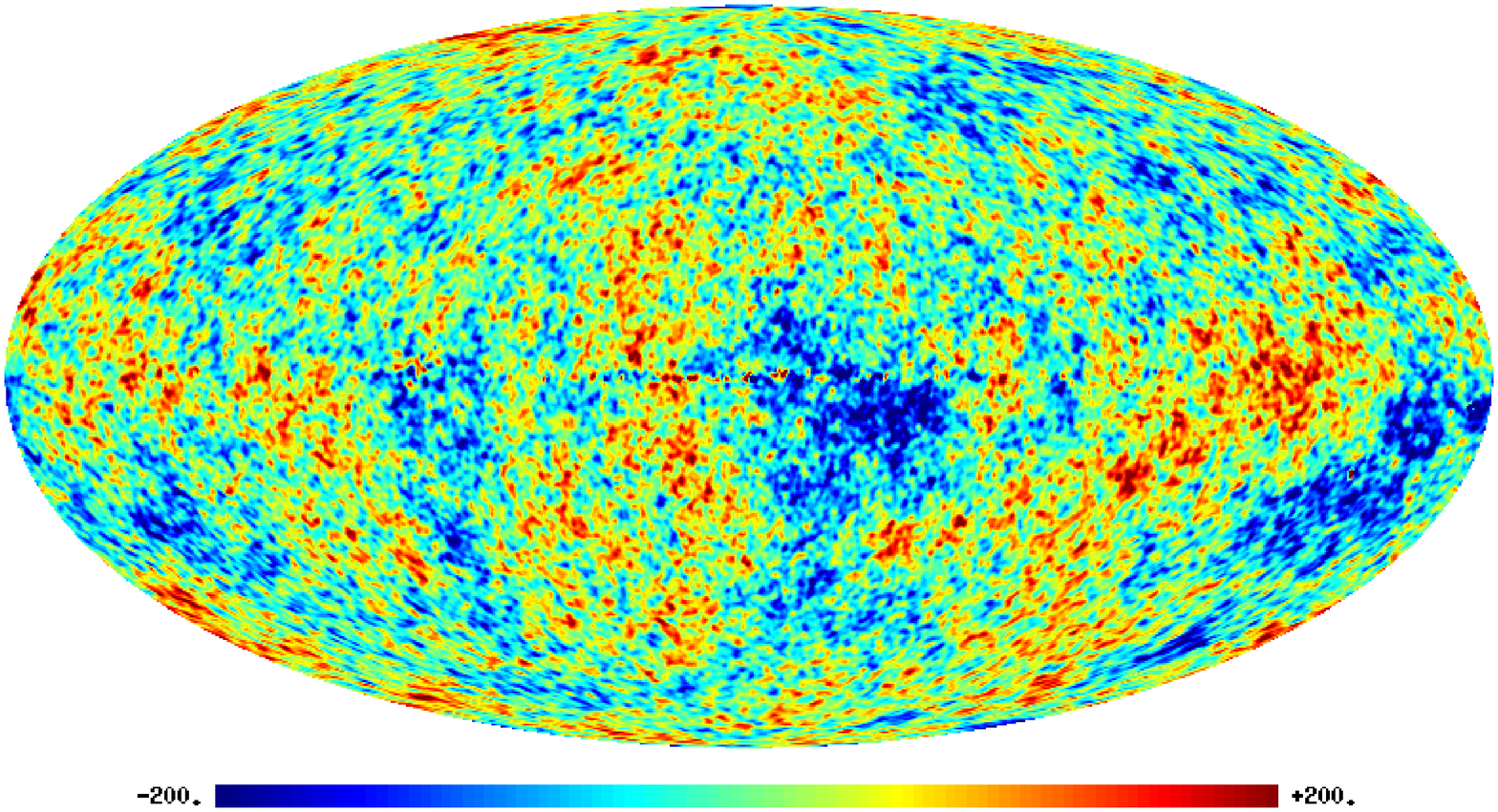}
\centering\includegraphics[scale=.25]{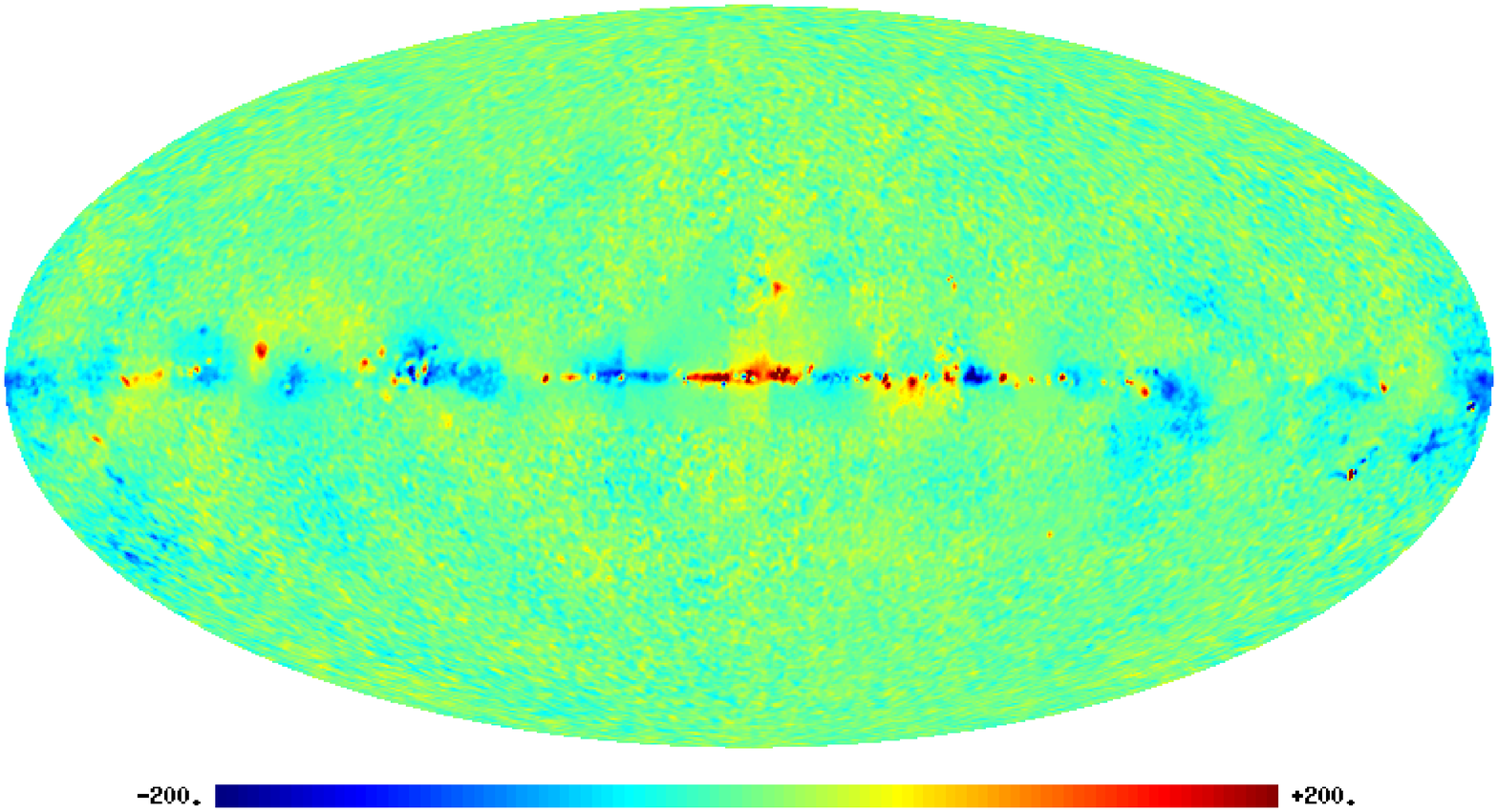}
\caption{HILC5YR (top), WILC5YR (middle), WILC5YR - HILC5YR (bottom)}
\label{HILC5YR2}
\end{figure}

We have also applied our method to the WMAP five year data \citep{WMAP:5yr_TT}, which have been released during the preparation of this paper. For comparison with HILC3YR, we have obtained HILC5YRa using the same HILC parameters as the HILC3YR (i.e. $l_{\mathrm{cutoff}}=7$ and $a^i_{lm}$ of the multipole range ($l\le 300$) for variance minimization). The HILC5YRa is shown with the HILC3YR in Fig. \ref{HILC5YR1}. 
We may see that the HILC5YRa contains less level of residual foreground, since most of the difference map shown in Fig. \ref{HILC5YR1} is positive. It is reported that some improvement in instrument calibration have been made for the WMAP 5 year data \citep{WMAP:5yr_TT}. We attribute less level of residual foregrounds in the HILC5YRa partially to the instrument calibration improvement, since it might improve the accuracy of frequency dependency of band map data.
We found that Eq. \ref{w_solution} used with the WMAP 5 year data has less degree of numerical singularity than the 3 year data, which may also be attributed to the improved accuracy of frequency dependency of band map data.
Since the WMAP 5 year data have higher SNR than the 3 year data, we have increased the multipole range of $a^i_{lm}$ to $l\le 400$.
The improved numerical stability and the increase in the multipole range of $a^i_{lm}$ ($l\le 400$) allowed us to
set $l_{\mathrm{cutoff}}$ to $15$. Hence we have obtained the HILC5YR with the $l_{\mathrm{cutoff}}=15$, which is shown with WILC5YR in Fig. \ref{HILC5YR2}.

\begin{figure}[htb!]
\includegraphics[scale=.55]{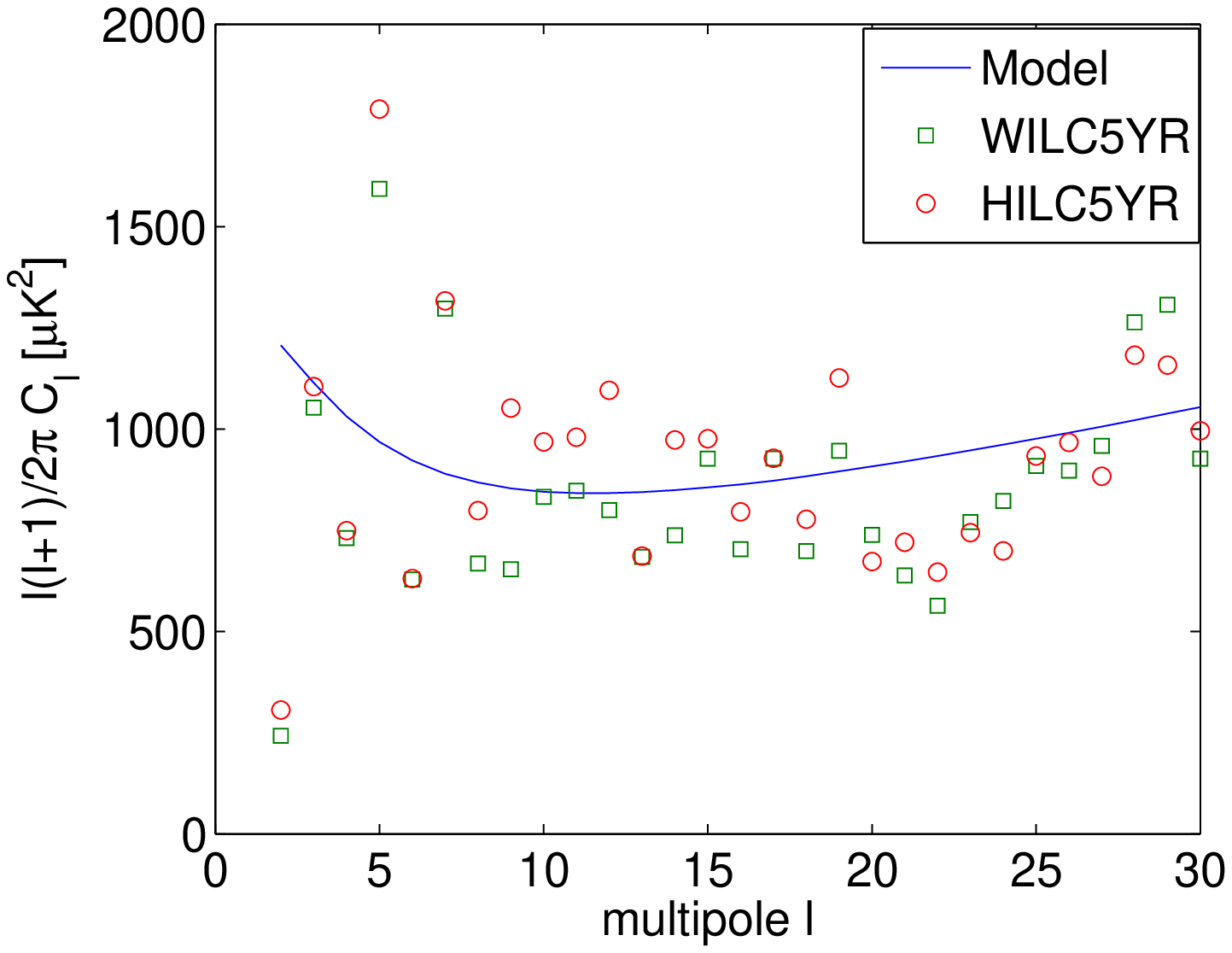}
\includegraphics[scale=.55]{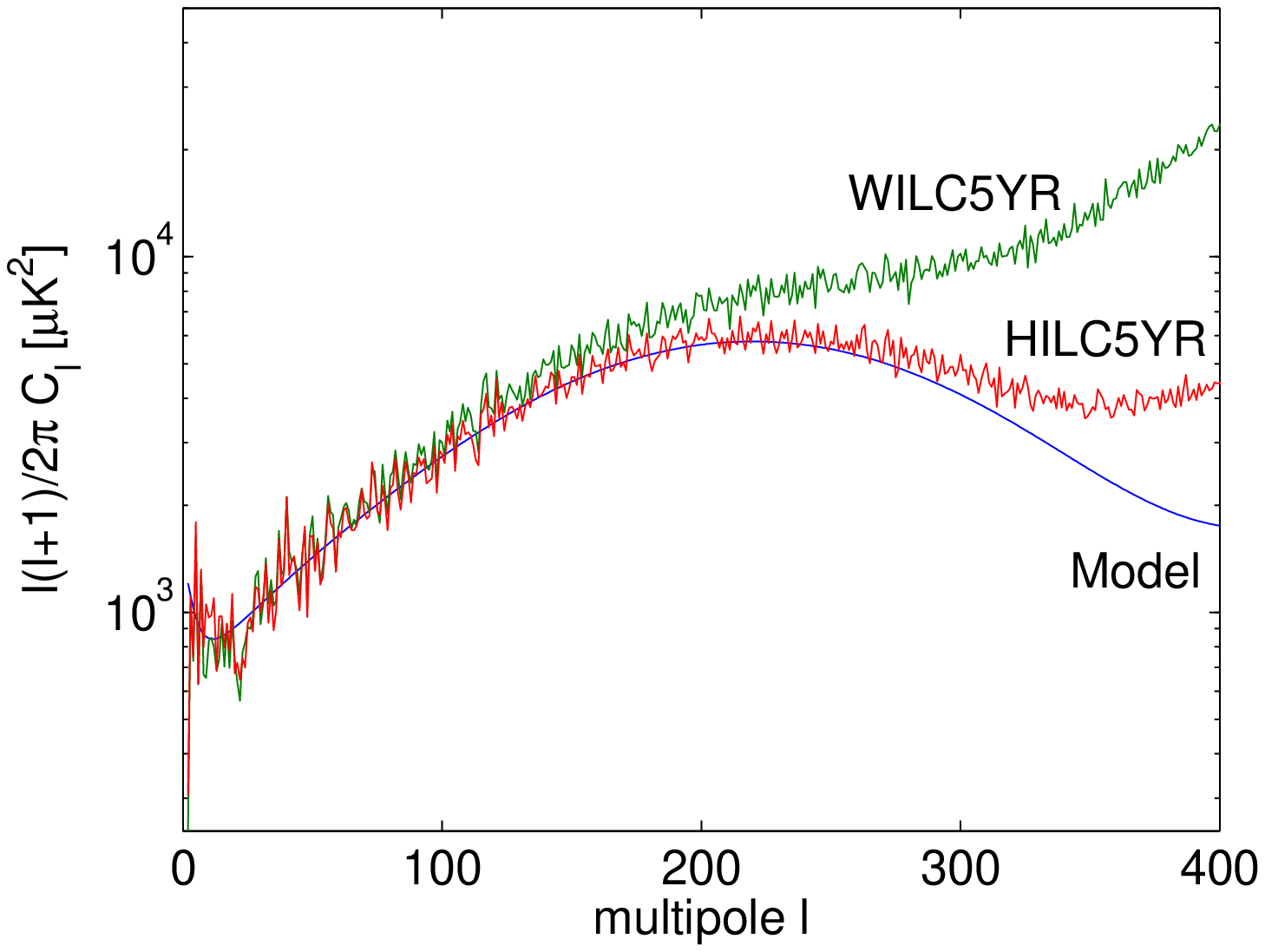}
\caption{Power spectra of HILC5YR, WILC5YR and the WMAP best-fit $\Lambda$CDM model}
\label{Cl_5yr}
\end{figure}

The power spectra estimate on HILC5YR and WILC3YR are shown with the WMAP 5 year best-fit $\Lambda$CDM model in Fig. \ref{Cl_5yr}.
By comparing Fig. \ref{Cl_5yr} with Fig. \ref{Cl_wmap}, we may see that noise level in 5 year ILC maps are lower than that of 3 year ILC maps, as expected.
It was reported that the existence of the cross term leads to the suppression of low multipole anisotropy of ILC maps \citep{Cosmic_Covariance,ILC_power}.
It is interesting to note that most of low multipole powers of WILC5YR are lower than those of HILC5YR.

\begin{table}[htb!]
\centering
\caption{quadrupole and octupole powers obtained with the WMAP 5 year data}
\begin{tabular}{cccc}
\hline\hline 
Measurement &  $\delta T^2_2$[$\mu\mathrm{K}^2$] & p-value &$\delta T^2_3$[$\mu\mathrm{K}^2$]\\ 
\hline
WMAP best-fit $\Lambda$CDM model &   1206.6& $\ldots$ & 1113.9\\ 
Hinshaw et al. cut sky &  213.4 & 2.86\%& 1038.7\\
WILC5YR & 242.7& 3.79\%& 1053.2\\
HILC5YR &  306.2 & 5.75\% & 1104.8\\ 
\hline 
\end{tabular}
\label{low_multipole_power2}
\end{table}
In Table \ref{low_multipole_power2}, the quadrupole and octupole powers of HILC5YR are shown with those of the WMAP 5 year best-fit $\Lambda$CDM model and WILC5YR. 

\begin{figure}[htb!]
\centering\includegraphics[scale=1]{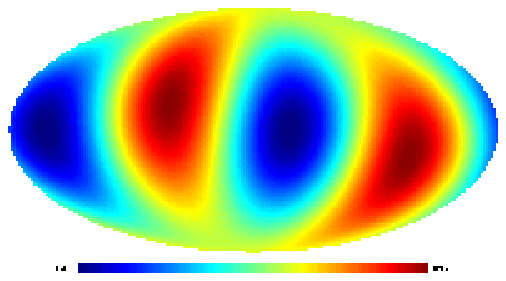}
\centering\includegraphics[scale=1]{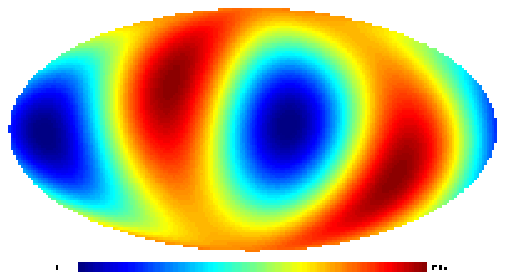}
\centering\includegraphics[scale=1]{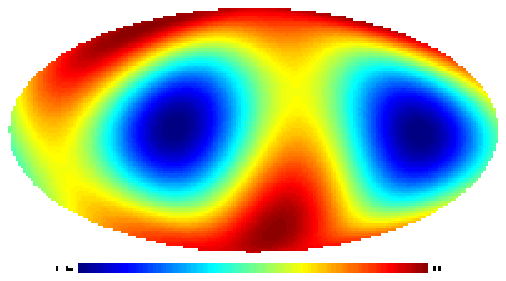}
\caption{Quadrupole Anisotropy [$\mu\mathrm K$]: HILC5YR (top), WILC5YR (middle), WILC5YR - HILC5YR (bottom)}
\label{ILC_l2}
\end{figure}

\begin{figure}[hbt!]
\centering\includegraphics[scale=1]{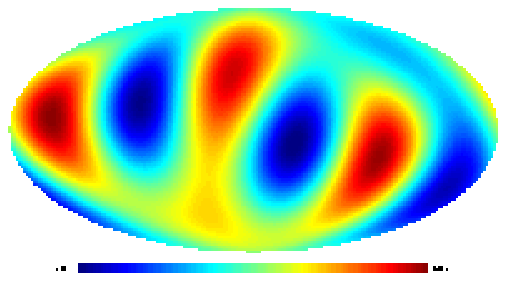}
\centering\includegraphics[scale=1]{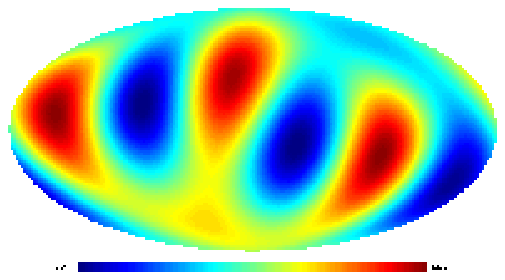}
\centering\includegraphics[scale=1]{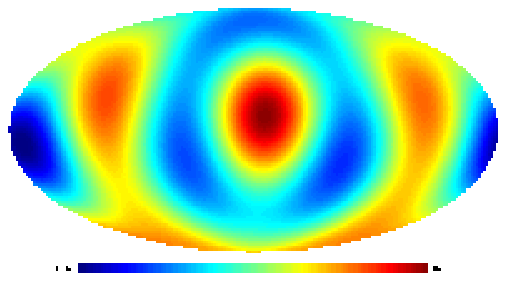}
\caption{Octupole Anisotropy [$\mu\mathrm K$]: HILC5YR (top), WILC5YR (middle), WILC5YR - HILC5YR (bottom)}
\label{ILC_l3}
\end{figure}

\begin{figure}[htb!]
\centering\includegraphics[scale=1]{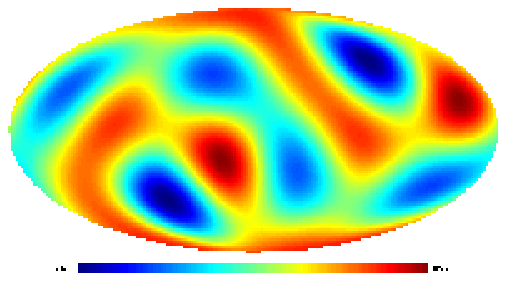}
\centering\includegraphics[scale=1]{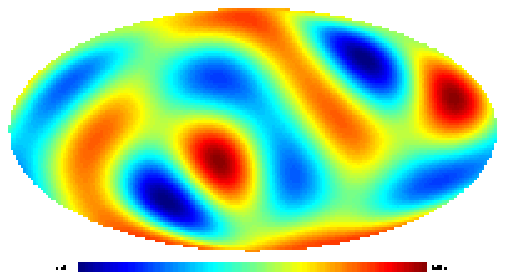}
\centering\includegraphics[scale=1]{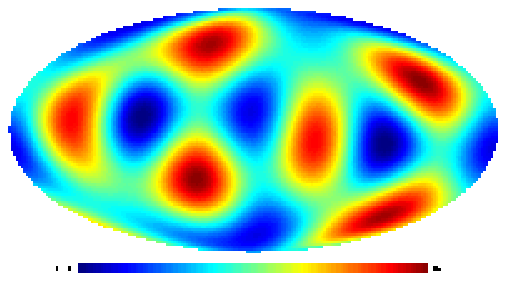}
\caption{Anisotropy of $l=4$ [$\mu\mathrm K$]: HILC5YR (top), WILC5YR (middle), WILC5YR - HILC5YR (bottom)}
\label{ILC_l4}
\end{figure}

\begin{figure}[htb!]
\centering\includegraphics[scale=1]{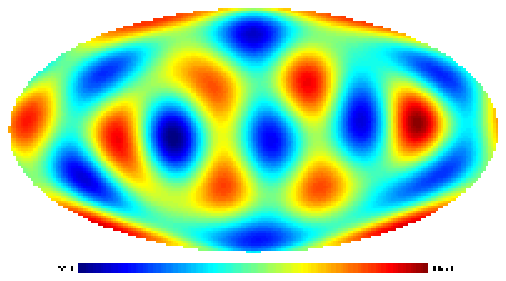}
\centering\includegraphics[scale=1]{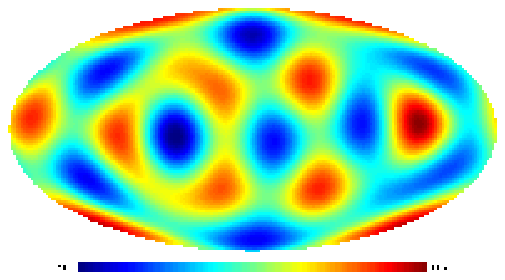}
\centering\includegraphics[scale=1]{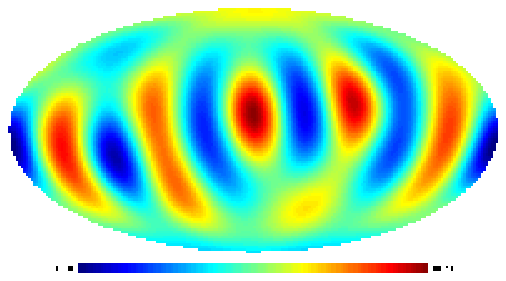}
\caption{Anisotropy of $l=5$: HILC5YR (top), WILC5YR (middle), WILC5YR - HILC5YR (bottom)}
\label{ILC_l5}
\end{figure}

The angular separation between the preferred axis of the quadrupole anisotropy and that of the octupole anisotropy is
$2^\circ$ for WILC5YR, while $12^\circ.1$ for HILC5YR. The corresponding probabilities of getting such an alignment
is $0.058\%$ for WILC5YR, while $2.17\%$ for HILC5YR.
The anisotropy of HILC5YR on low multipoles ($2\le l \le 5$) are shown in Fig. \ref{ILC_l2}, \ref{ILC_l3}, \ref{ILC_l4} and \ref{ILC_l5} with those of WILC5YR and difference maps. 

\begin{figure}[htb!]
\centering\includegraphics[scale=1]{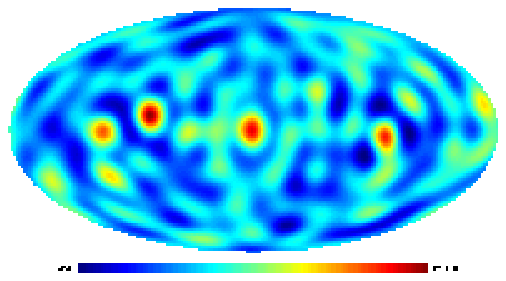}
\centering\includegraphics[scale=1]{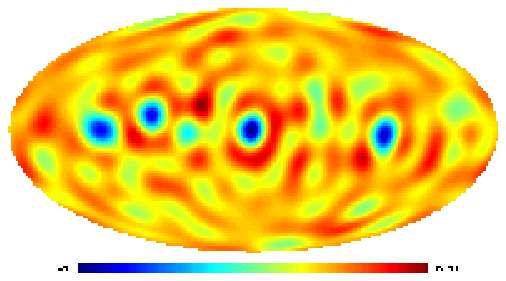}
\centering\includegraphics[scale=1]{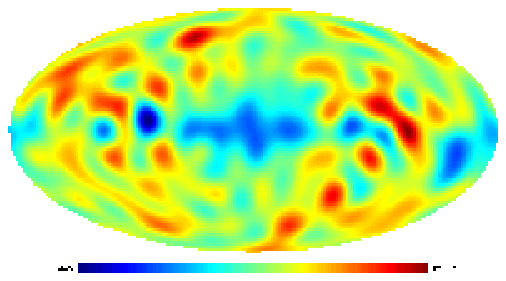}
\centering\includegraphics[scale=1]{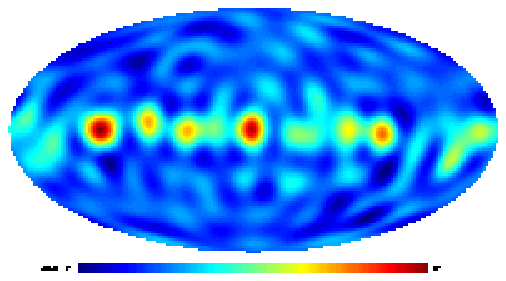}
\centering\includegraphics[scale=1]{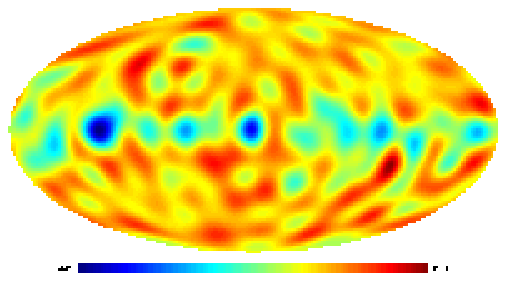}
\caption{The HILC5YR linear weights for the K, Ka, Q, V, and W band map (from top to bottom)}
\label{W_HILC5YR2}
\end{figure}
\begin{figure}[htb!]
\centering\includegraphics[scale=.52]{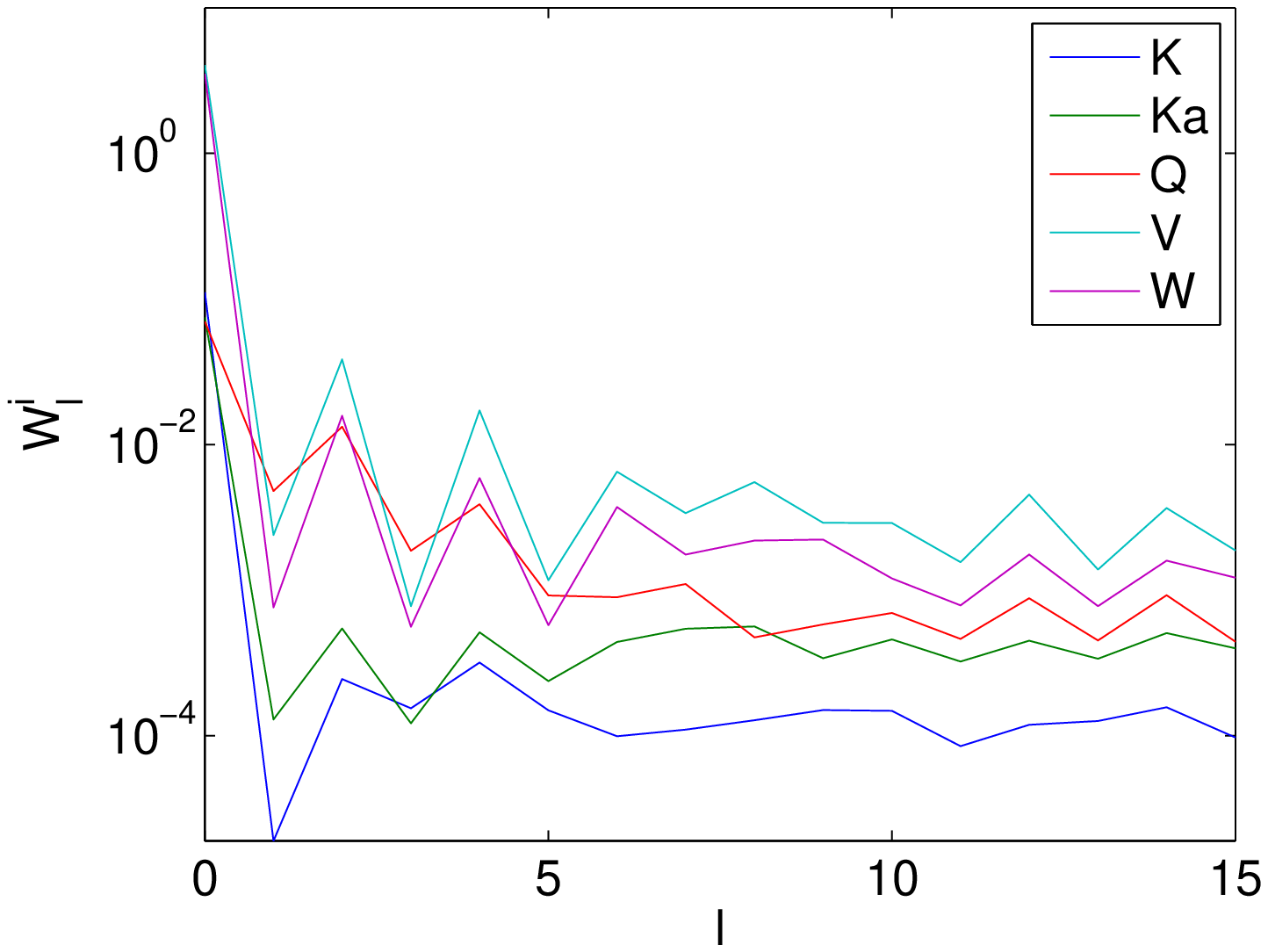}
\caption{Variance of HILC5YR linear weights}
\label{Wl_5YR}
\end{figure}

The linear weights of HILC5YR are shown in Fig. \ref{W_HILC5YR2}.
In Fig. \ref{Wl_5YR}, we show the variance of the linear weights of HILC5YR, which is computed by $W^i_l=(2l+1)^{-1}\sum_m |w^i_{lm}|^2$.

\section{Computational Issue}
\label{computation}
The computation of linear weights by our harmonic variance minimization can be split as follows:\\
1) Computing $\gamma_i(l_1,m_1,l_3,m_3)$ by Eq. \ref{gamma}\\
2) Computing $\alpha^{i'i}_{l'm'lm}$ by Eq. \ref{alpha}\\
3) Solving the system of linear equations given by Eq. \ref{minimum_wlm}.

Let's assume that there are $f_n$ band maps of high SNR up to some multipole $L$
and the cutoff multipole for linear weights is set to $l$. 
With the recurrence relation \citep{Wigner3j}, we are at present able to compute Wigner 3j symbols in Eq. \ref{gamma} up to high multipoles ($\sim 700$) fast enough. Therefore, step 1) put relatively little computational load.\\
Step 2) requires $\mathcal O(\mathcal N^2\,\mathcal M)$ and  step 3), which involves $\mathcal N$ by $\mathcal N$ matrix inversion, requires  $\mathcal O(\mathcal N^3)$, where $\mathcal N=(l+1)^2\,f_n$ and $\mathcal M=(L+1)^2$.
$l$ is chosen to be much smaller than $L$ to keep matrices in Eq. \ref{w_solution} numerically non-singular (e.g. In the application to WMAP data, we have set $l=7$ and $L=300$, with consideration of numerical singularity and the WMAP band map's SNR.) Since  $\mathcal N=(l+1)^2\,f_n$ is much smaller than $\mathcal M=(L+1)^2$, the total computing time is $O(\mathcal N^2\,\mathcal M)$.

As described previously, we have found $l_\mathrm{cutoff}$ phenomenologically by increasing it until numerical singularity in Eq. \ref{w_solution} emerges. While ideally $l_\mathrm{cutoff}$ can be as high as $L$, the optimal value of $l_\mathrm{cutoff}$ for the WMAP data seems to be much smaller than $L$. 
The numerical singularity for ($l_\mathrm{cutoff}<L$) may be attributed to large bandwidth and relatively small separation of the WMAP frequency channels, because in such configurations the frequency spectrum of foregrounds may not be numerically distinct enough over the channels. Through simple extrapolation by comparing the beamwidth and SNR of the Planck surveyor with those of the WMAP, we may make rough estimate that HILC method with $l_\mathrm{cutoff}> 100$ may be numerically stable for the Planck temperature data.

\section{Conclusion}
\label{conclusion}
In spite of the warning from the WMAP team against serious use of the ILC map, the ILC map has been widely used especially for
low multipole anisotropy study, since the template-subtraction maps are not suitable for a whole sky map due to heavy foreground contamination within Kp2 cut. We have summarized the advantages and disadvantages of ILC variants and the template-fitting method in appendix \ref{comparison}.

For these reason, we have pursued the ILC method and extended it by improving the causes related to the complication in noise properties (e.g.  smoothing on the disjoint regions, Monte-Carlo `bias' correction).  Through our effort toward improved ILC implementation, we have developed a harmonic variance minimization method, which is derived by converting a general form of pixel-domain approach into spherical harmonic space. 
In our approach, spatial variability of linear weights is incorporated in a self-contained manner and linear weights are continuous over whole sky.
Thanks to full correspondence to a general pixel-domain method, physical interpretation on our method is quite straightforward.
In variance minimization, there exists a cross term between residual foreground and CMB, which makes the linear combination of minimum variance differ from that of minimum foreground.
We have developed an iterative method, where perturbative correction is made for the cross term.

The simulations showed that our method yields reliable and stable reconstruction of the CMB also on lowest multipoles.
By applying it to the WMAP data, we have obtained a CMB map, whose power spectra makes better agreement with the WMAP best-fit $\Lambda$ CDM model than the WILC5YR. 
The CMB map and linear weights, which we have obtained, are available from \texttt{http://www.nbi.dk/$\sim$jkim/hilc}.

Capacity of our method to clean foreground, whose spectral indice is a rapidly varying function of position, is limited by the SNR of data on high multipoles, since increasing the cutoff multipole of linear weights requires more data to avoid numerical singularity in Eq. \ref{w_solution}. When the data of good SNR on high multipoles are available, we will be able to make better reduction of foregrounds by setting the cutoff multipole to higher value. 
This is similar to saying that we are able to assume finer disjoint regions in pixel-domain approach, when data of better pixel resolution are available.

The foreground reduction method by template fitting is unable to take full advantage of the high resolution low noise Planck data, since the available templates do not have such high resolution and low noise over a whole sky. 
Unlike the template method, the effectiveness of our method scales with frequency channel numbers and 
angular resolution of the observation data. Hence our method is more suitable for the Planck surveyor, which has nine frequency channels with good SNR and angular resolution.

Our method, which is presented for the application to temperature data, may be easily extended to the polarization data by making the following replacement in Eq. \ref{gamma}:
\[\left(\begin{array}{ccc}l_1&l_2&l_3\\0&0&0\end{array}\right)\;\;\;\rightarrow\;\;\;\left(\begin{array}{ccc}l_1&l_2&l_3\\0&\pm2&\mp2\end{array}\right).\] 
We believe a blind approach like our method is desirable for the analysis of polarization data from the upcoming Planck satellite, since the availability of polarized foreground templates is quite limited. 
For these reasons, the method presented in this paper will stand out among several foreground reduction methods, when low-noise high resolution data of CMB temperature and polarization are available from the Planck surveyor \citep{Planck:sensitivity}.

\section{ACKNOWLEDGMENTS}
We are grateful to Changbom Park for helpful discussions.
We acknowledge the use of the Legacy Archive for Microwave Background Data Analysis (LAMBDA). 
Our simulation and data analysis made the use of HEALPix\cite{HEALPix:Primer,HEALPix:framework} and
were performed on the supercomputing facility of the Danish Center for Scientific Computing.
This work was supported by FNU grant 272-06-0417, 272-07-0528 and 21-04-0355. 
\begin{appendix}
\section{Computing linear weights via matrix operations}
\label{matrix_solution}
We present the solutions of Eq. \ref{w00_constraint}, \ref{wlm_constraint} and \ref{minimum_wlm} through matrix operations.
Consider $n$ linear weights for maps of $k$ frequency channels.
In matrix notation, the constraints given by Eq. \ref{w00_constraint} and \ref{wlm_constraint} are as follows:
\begin{eqnarray}
\mathbf \Pi\cdot \mathbf w&=&\mathbf {e} \label{constraint}.
\end{eqnarray}
$\mathbf \Pi$ is a $\frac{n}{k}\times n$ matrix, given by
\begin{eqnarray*}
\mathbf \Pi_{ij}=\
&&\left\{\begin{array}{r@{\quad:\quad}l}
1&k(i-1)+1\le j \le k\,i\\
0& \mathrm {otherwise}\end{array}\right.
\end{eqnarray*}
and $\mathbf e$ is a column vector of length $n/k$, given by
\begin{eqnarray*}
\mathbf e_{j}=
&&\left\{\begin{array}{r@{\quad:\quad}l}
\sqrt{4\pi}&j=1\\0&
j>1 \end{array}\right.
\end{eqnarray*}
In matrix notation, Eq. \ref{minimum_wlm} is as follows:
\begin{eqnarray}
\mathbf A \cdot \mathbf w&=&-\mathbf{\Pi}^{\mathrm{T}}\mathbf{L},\label{AwL}
\end{eqnarray}
where
\begin{eqnarray*}
\mathbf A_{j'j}&=&\alpha^{i'i}_{l'm'lm},\\
\mathbf w_{j}&=&\tilde{w}^i_{lm},\\
\mathbf L_{j''}&=&\lambda_{l'm'},
\end{eqnarray*}
for $j'=k(l'^2+l'+m')+i'$, $j=k(l^2+l+m)+i$, and $j''=l'^2+l'+m'$.
$\mathbf A$ is a $n\times n$ matrix, and $\mathbf w$ and $\mathbf L$ are column vectors of length $n$, and 
length $n/k$ respectively.
With Eq. \ref{AwL}, $\mathbf w$ is solved in terms of $n/k$ undetermined Langrange multipliers, provided that $\mathbf A$ is invertible:
\begin{eqnarray}
\mathbf w=-\mathbf A^{-1}\mathbf{\Pi}^{\mathrm{T}}\,\mathbf{L}.\label{wAL}
\end{eqnarray}
With Eq. \ref{constraint} and \ref{wAL}, the undetermined $n/k$ Langrange multipliers are given by:
\begin{eqnarray}
\mathbf L=-(\mathbf{\Pi}\,\mathbf A^{-1}\,\mathbf{\Pi}^{\mathrm{T}})^{-1} \mathbf e,\label{Lagrange}
\end{eqnarray}
Therefore, linear weights in spherical harmonic space is given as follows:
\begin{eqnarray}
\mathbf w=\mathbf A^{-1}\mathbf{\Pi}^{\mathrm{T}}(\mathbf{\Pi}\,\mathbf A^{-1}\,\mathbf{\Pi}^{\mathrm{T}})^{-1} \mathbf e.\label{w_solution}
\end{eqnarray}
Eq. \ref{w_solution} is not reduced to $\mathbf w=\mathbf{\Pi}^{-1} \mathbf e$, since $\mathbf{\Pi}$ and $\mathbf{\Pi}^\mathrm{T}$ are not square matrices.

\section {The effect of the cross term in variance minimization}
\label{supprression}
To better understand the effect of the cross term, we investigate a simple case of two frequency observation channels and foregrounds of uniform frequency spectra. 
Since the analysis in Sec. \ref{simulation} and \ref{WMAP} are carried out on the multipole of high SNR, 
we neglect instrument noise.
We assume linear weights to be constant, because the foreground spectra are assumed to be spatially uniform.
Keeping the CMB signal unchanged, we assign constant linear weights $w$ and $(1-w)$ for the frequency channel 1 and 2 respectively.
Since spherical harmonic coefficients of the linear combination map is
\[a_{lm}=w\,a^{\mathrm{fg1}}_{lm}+(1-w)\,a^{\mathrm{fg2}}_{lm}+a^{\mathrm{cmb}}_{lm},\]
the variance of the linear combination map is as follows:
\begin{eqnarray}
\sigma^2=\sum_{l'm'}\left|w\,a^{\mathrm{fg1}}_{l'm'}+(1-w)\,a^{\mathrm{fg2}}_{l'm'}+a^{\mathrm{cmb}}_{l'm'}\right|^2,\label{sigma}
\end{eqnarray}
where $a^{\mathrm{fg1}}_{l'm'}$ and $a^{\mathrm{fg2}}_{l'm'}$ denote the spherical harmonic coefficients of foregrounds at frequency channel 1 and 2 respectively.
Since the value of $w$, which minimizes Eq. \ref{sigma}, is 
\begin{eqnarray*}
\frac{-\sum\limits_{l'm'}\mathrm{Re}\left[(a^{\mathrm{fg1}}_{l'm'}-a^{\mathrm{fg2}}_{l'm'})\left(a^{\mathrm{fg2}}_{l'm'}+a^{\mathrm{cmb}}_{l'm'}\right)^* \right]}{\sum\limits_{l'm'} |a^{\mathrm{fg1}}_{l'm'}-a^{\mathrm{fg2}}_{l'm'}|^2},
\end{eqnarray*}
the linear combination map of minimum variance has the following spherical harmonic coefficients: 
\begin{eqnarray*}
\lefteqn{a_{lm}=a^{\mathrm{cmb}}_{lm}+
\frac{1}{\sum\limits_{l'm'} |a^{\mathrm{fg1}}_{l'm'}-a^{\mathrm{fg2}}_{l'm'}|^2}
\left(a^{\mathrm{fg2}}_{lm}\sum\limits_{l'm'} |a^{\mathrm{fg1}}_{l'm'}-a^{\mathrm{fg2}}_{l'm'}|^2\right.}\nonumber\\
&&\left.-(a^{\mathrm{fg1}}_{lm}-a^{\mathrm{fg2}}_{lm})\sum\limits_{l'm'}\mathrm{Re}\left[(a^{\mathrm{fg1}}_{l'm'}-a^{\mathrm{fg2}}_{l'm'})\left(a^{\mathrm{fg2}}_{l'm'}+a^{\mathrm{cmb}}_{l'm'}\right)^* \right]\right).\nonumber\\
\end{eqnarray*}
The power spectra of our galactic foregrounds has uneven distribution with high concentration on lowest multipole and azimuthal mode
(i.e. $a^{\mathrm{fg}}_{20}\gg a^{\mathrm{fg}}_{l'm'}$). Therefore, the spherical harmonic coefficients of the linear combination map is 
\begin{eqnarray}
a_{lm}\approx a^{\mathrm{cmb}}_{lm}+a^{\mathrm{fg2}}_{lm}
-(a^{\mathrm{fg1}}_{lm}-a^{\mathrm{fg2}}_{lm})\frac{a^{\mathrm{fg2}}_{20}+a^{\mathrm{cmb}}_{20}}{a^{\mathrm{fg1}}_{20}-a^{\mathrm{fg2}}_{20}}.\label{alm_two_channel_approx}
\end{eqnarray}
According to Eq. \ref{alm_two_channel_approx}, $a_{lm}\approx0$ for ($l=2,m=0$). This is in agreement with the suppression on lowest multipole powers in an Internal Linear Combination (ILC) map, reported by \citep{ILC_power,Cosmic_Covariance}.
Therefore, we attribute the existence of the cross term and highly uneven power spectrum of foregrounds to the suppression on low multipole power in an Internal Linear Combination (ILC) map.
To test our hypothesis, we have carried out simulations with foregrounds of flat power spectrum, which are derived from the WMAP Maximum-Entropy-Method(MEM) foregrounds as follows:
\begin{eqnarray}
a^{\mathrm{fg}'}_{lm}=\sqrt{\frac{\sum^{l_1}_{l_0}C^{\mathrm{fg}}_l}{l_1-l_0+1}}\frac{a^{\mathrm{fg}}_{lm}}{\sqrt(C^{\mathrm{fg}}_l)}, 
\end{eqnarray}
where $a^{\mathrm{fg}}_{lm}$ is spherical harmonic coefficients of the WMAP MEM foreground and $C^{\mathrm{fg}}_l=\sum_m |a^{\mathrm{fg}}_{lm}|^2/(2l+1)$.
The role of $\sqrt{\sum_{l} C^{\mathrm{fg}}_l/(l_1-l_0+1)}$ is to match the total power of the flat foreground with that of the MEM foregrounds.
The result of simulations show that there is no observable suppression on low multipole powers when foregrounds have flat power spectrum, thereby supporting our hypothesis.
 
This suppression on low multipole power can be reduced by excluding the multipoles of high foreground concentration (i.e. low multipoles) in variance minimization process. Consider minimizing the following variance where low multipoles are excluded:
\begin{eqnarray*}
\sigma^2=\sum^{l_1}_{l'>l_0}\left|w\,a^{\mathrm{fg1}}_{l'm'}+(1-w)\,a^{\mathrm{fg2}}_{l'm'}+a^{\mathrm{cmb}}_{l'm'}\right|^2,
\end{eqnarray*}
where foreground power in the multipole range ($l'\le l_0$) are much bigger than those on higher multipoles. 
Then, the linear combination map has the following spherical harmonic coefficients:
\begin{eqnarray}
\lefteqn{a_{lm}=a^{\mathrm{cmb}}_{lm}+
\frac{1}{\sum\limits_{l'>l_0,m'} |a^{\mathrm{fg1}}_{l'm'}-a^{\mathrm{fg2}}_{l'm'}|^2}}\label{alm_no_bias}\\
&&\times\left(a^{\mathrm{fg2}}_{lm}\sum\limits_{l'>l_0,m'} |a^{\mathrm{fg1}}_{l'm'}-a^{\mathrm{fg2}}_{l'm'}|^2-(a^{\mathrm{fg1}}_{lm}-a^{\mathrm{fg2}}_{lm})\right.\nonumber\\
&&\left.\times\sum\limits_{l'>l_0,m'}\mathrm{Re}\left[(a^{\mathrm{fg1}}_{l'm'}-a^{\mathrm{fg2}}_{l'm'})\left(a^{\mathrm{fg2}}_{l'm'}+a^{\mathrm{cmb}}_{l'm'}\right)^* \right]\right).\nonumber
\end{eqnarray}
With enough number of summation terms,  the term inside the big parenthesis of Eq. \ref{alm_no_bias} approaches zero because of cancellation. If we had not excluded low multipoles, the cancellation would be ineffective, due to asymmetric distribution of foreground power (i.e. high concentration on low multipoles.).
Eq. \ref{alm_no_bias} is valid for $a_{lm}$ ($l\le l_0$) as well as ($l>l_0$).
In other words, though the linear weights were determined through variance minimization over ($l_0<l'$), they are applicable to foreground reduction in $a_{lm}$ ($l\le l_0$).

We have also investigated the non-uniform frequency spectra case by resorting to a numerical investigation. We find that there is a tendency of suppression over multipole range ($l_0-l_{\mathrm{cutoff}}<l<l_0+l_{\mathrm{cutoff}}$), where $l_{\mathrm{cutoff}}$ is the assumed cutoff multipole of $w^i_{lm}$.
We also find that the suppression is reduced by excluding the multipoles of high foreground concentration in variance minimization process, just as in the uniform frequency spectra case.\\

\section{Comparison of the ILC method variants and the template-fitting method}
\label{comparison}
In this section, we are discussing briefly the advantages and disadvantages of ILC variants and a template fitting method.
The template fitting method is the foreground reduction method most importantly employed by the WMAP team. It has advantage that it has less complicated noise properties and is free from the Cosmic Covariance problem, while it has the disadvantages that it relies on foreground templates of external sources, hence requiring extrapolation to observation frequencies and currently unable to provide a whole sky map, due to heavy foreground contamination within Kp2 cut.
WILC is the ILC implementation by the WMAP team. 
It has the advantages that it utilizes the boundary shape information of galactic foregrounds and scales with the number of observation frequency channels, while its disadvantages are sharp boundaries of disjoint regions (Smoothing boundaries in the final map making does not solve discontinuity problem completely, since region definition with sharp boundary are used in variance minimization.), dubious bias correction of the `Cosmic Covariance' by Monte-Carlo CMB, its reliance on the pre-defined disjoint regions (not being a completely blind approach).
TCM3YR \citep{Tegmark:CMB_map} is the ILC variant, where the variance of each multipole is minimized separately.
It has the advantages that the dependency of foreground power on angular scales is reflected and scales  with the number of frequency channels, while it has  the disadvantages of sharp boundaries, `Cosmic Covariance' problem, 
need for the pre-defined disjoint regions.
SILC3YR \citep{Park:SILC400} is another ILC variant. Instead of using disjoint region definition by the WMAP team, disjoint regions were derived from the MEM reconstructed foregrounds.
It has the advantages that the definition of the disjoint regions may be optimal than those of WILC, and it scales with the number of frequency channels, while it has the disadvantages of sharp boundaries, `Cosmic Covariance' problem, need for the MEM foreground (traces MEM foregrounds and hence WILC3YR).
HILC, which is our ILC implementation, has the advantages that it does not rely on the definition of disjoint regions (hence no sharp boundaries), scalability with the number of observation frequency channels and angular resolution of data, while it has the disadvantages that it is computationally intensive, and has the `Cosmic Covariance' problem, though reduced.
\end{appendix}


\begin{thebibliography}{0}
\expandafter\ifx\csname natexlab\endcsname\relax\def\natexlab#1{#1}\fi
\expandafter\ifx\csname bibnamefont\endcsname\relax
  \def\bibnamefont#1{#1}\fi
\expandafter\ifx\csname bibfnamefont\endcsname\relax
  \def\bibfnamefont#1{#1}\fi
\expandafter\ifx\csname citenamefont\endcsname\relax
  \def\citenamefont#1{#1}\fi
\expandafter\ifx\csname url\endcsname\relax
  \def\url#1{\texttt{#1}}\fi
\expandafter\ifx\csname urlprefix\endcsname\relax\def\urlprefix{URL }\fi
\providecommand{\bibinfo}[2]{#2}
\providecommand{\eprint}[2][]{\url{#2}}

\end{thebibliography}


\begin{thebibliography}{16}
\expandafter\ifx\csname natexlab\endcsname\relax\def\natexlab#1{#1}\fi
\expandafter\ifx\csname url\endcsname\relax
  \def\url#1{{\tt #1}}\fi

\bibitem[Bennett et~al.(2003)Bennett, Hill, Hinshaw, Nolta, Odegard, Page,
  Spergel, Weiland, Wright, Halpern, Jarosik, Kogut, Limon, Meyer, Tucker, and
  Wollack]{WMAP:foreground}
C.~Bennett, R.~S. Hill, G.~Hinshaw, M.~R. Nolta, N.~Odegard, L.~Page, D.~N.
  Spergel, J.~L. Weiland, E.~L. Wright, M.~Halpern, N.~Jarosik, A.~Kogut,
  M.~Limon, S.~S. Meyer, G.~S. Tucker, and E.~Wollack.
\newblock First year wilkinson microwave anisotropy probe ({WMAP})
  observations: Foreground emission.
\newblock {\em Astrophys.J.Suppl.}, 148, 97, 2003.
\newblock http://lambda.gsfc.nasa.gov.

\bibitem[Hinshaw et~al.(2003)Hinshaw, Spergel, Verde, Hill, Meyer, Barnes,
  Bennett, Halpern, Jarosik, Kogut, Komatsu, Limon, Page, Tucker, Weiland,
  Wollack, and Wright]{WMAP:TT}
G.~Hinshaw, D.~N. Spergel, L.~Verde, R.~S. Hill, S.~S. Meyer, C.~Barnes, C.~L.
  Bennett, M.~Halpern, N.~Jarosik, A.~Kogut, E.~Komatsu, M.~Limon, L.~Page,
  G.~S. Tucker, J.~Weiland, E.~Wollack, and E.~L. Wright.
\newblock First year {Wilkinson Microwave Anisotropy Probe} ({WMAP})
  observations: Angular power spectrum.
\newblock {\em Astrophys.J.Suppl.}, 148,  135, 2003.

\bibitem[Hinshaw and et~al.(2007)]{WMAP:3yr_TT}
G.~Hinshaw and et~al.
\newblock Three-year {Wilkinson Microwave Anisotropy Probe} ({WMAP})
  observations: Temperature analysis.
\newblock {\em Astrophys.J.Suppl.}, 170,  288, 2007.
\newblock http://lambda.gsfc.nasa.gov.

\bibitem[de~Oliveira-Costa and Tegmark(2006)de~Oliveira-Costa, and Tegmark]{Tegmark:foreground_presence}
Angelica de~Oliveira-Costa and Max Tegmark.
\newblock {CMB} multipole measurements in the presence of foregrounds.
\newblock {\em Phys. Rev. D}, 74, 023005, 2006.

\bibitem[Eriksen et~al.(2004)Eriksen, Banday, Gorski, and Lilje]{Eriksen:ILC}
H.~K. Eriksen, A.~J. Banday, K.~M. Gorski, and P.~B. Lilje.
\newblock On foreground removal from the {W}ilkinson {M}icrowave {A}nisotropy
  {P}robe data by an {I}nternal {L}inear {C}ombination method: Limitations and
  implications.
\newblock {\em Astrophys. J.}, 612,  633, 2004.

\bibitem[Davies et~al.(2006)Davies, Dickinson, Banday, Jaffe, and
  Gorski]{Foreground:Spectra}
R.~D. Davies, C.~Dickinson, A.J. Banday, T.~R. Jaffe, and K.~M. Gorski.
\newblock A determination of the spectra of galactic components observed by
  wmap.
\newblock {\em Mon. Not. R. Astron. Soc.}, 370, 1125, 2006.

\bibitem[Tegmark, de~Oliveira-Costa and Hamilton(2003)Tegmark, de~Oliveira-Costa, and
  Hamilton]{Tegmark:CMB_map}
M. Tegmark, A. de~Oliveira-Costa, and A. Hamilton.
\newblock A high resolution foreground cleaned {CMB} map from {WMAP}.
\newblock {\em Phys. Rev. D}, 68, 123523, (2003).


\bibitem[Arfken and Weber(2000)]{Arfken}
George~B. Arfken and Hans~J. Weber.
\newblock {\em Mathematical Methods for Physicists}.
\newblock Academic Press, San Diego, CA USA, 5th edition, 2000.


\bibitem[Chiang et~al.(2007)Chiang, Naselsky, and Coles]{Cosmic_Covariance}
Lung-Yih Chiang, Pavel~D. Naselsky, and Peter Coles.      

\newblock Cosmic covariance and the low quadrupole anisotropy of the Wilkinson
  Microwave Anisotropy Probe {WMAP} data.
\newblock {\em submitted to ApJ},
\newblock arXiv:0711.1860.

\bibitem[Tegmark and Efstathiou(1996)]{Tegmark:Foreground}
M.~Tegmark and G.~Efstathiou.
\newblock A method for subtracting foregrounds from multi-frequency {CMB} sky
  maps.
\newblock {\em Mon. Not. R. Astron. Soc.}, 281,  1297, 1996.


\bibitem[de~Oliveira-Costa et~al.(2004)de~Oliveira-Costa, Tegmark, Zaldarriaga,
  and Hamilton]{Tegmark:Alignment}
Angelica de~Oliveira-Costa, Max Tegmark, Matias Zaldarriaga, and Andrew
  Hamilton.
\newblock The significance of the largest scale {CMB} fluctuations in {WMAP}.
\newblock {\em Phys. Rev. D}, 69:063516, 2004.

\bibitem[Hinshaw et~al.()Hinshaw, Weiland, Hill, Odegard, Larson, Bennett,
  Dunkley, Gold, Greason, Jarosik, Komatsu, Nolta, Page, Spergel, Wollack,
  Halpern, Kogut, Limon, Meyer, Tucker, and Wright]{WMAP:5yr_TT}
G.~Hinshaw, J.~L. Weiland, R.~S. Hill, N.~Odegard, D.~Larson, C.~L. Bennett,
  J.~Dunkley, B.~Gold, M.~R. Greason, N.~Jarosik, E.~Komatsu, M.~R. Nolta,
  L.~Page, D.~N. Spergel, E.~Wollack, M.~Halpern, A.~Kogut, M.~Limon, S.~S.
  Meyer, G.~S. Tucker, and E.~L. Wright.
\newblock Five-year wilkinson microwave anisotropy probe (wmap) observations:
  Data processing, sky maps, and basic results.

\bibitem[Schulten and Gordon(1975)]{Wigner3j}
Klaus Schulten and Roy~G. Gordon.
\newblock Exact recursive evaluation of 3-j and 6-j coefficients for
  quantum-mechanical coupling of angular momenta.
\newblock {\em J. Math. Phys}, 16,  1961, 1975.


\bibitem[Gorski et~al.(2005)Gorski, Hivon, Banday, Wandelt, Hansen, Reinecke,
  and Bartelman]{HEALPix:framework}
K.~M. Gorski, E.~Hivon, A.~J. Banday, B.~D. Wandelt, F.~K. Hansen, M.~Reinecke, and M.~Bartelman.
\newblock {HEALPix} -- a framework for high resolution discretization, and fast
  analysis of data distributed on the sphere.
\newblock {\em Astrophys. J.}, 622,  759, 2005.

\bibitem[Gorski et~al.(1999)Gorski, Wandelt, Hansen, Hivon, and
  Banday]{HEALPix:Primer}
Krzysztof~M Gorski, Benjamin~D. Wandelt, Frode~K. Hansen, Eric Hivon, and
  Anthony~J. Banday.
\newblock The {HEALPix} primer.
\newblock astro-ph/9905275, 1999.

\bibitem[Saha et~al.(2007)Saha, Prunet, Jain, and Souradeep]{ILC_power}
Rajib Saha, Simon Prunet, Pankaj Jain, and Tarun Souradeep.
\newblock {CMB} anisotropy power spectrum using linear combinations of {WMAP}
  maps.
\newblock arXiv:0706.3567, 2007.


\bibitem[Tauber(2000)]{Planck:sensitivity}
J.~A. Tauber.
\newblock The Planck mission: Overview and current status.
\newblock {\em Astrophysical Letters and Communications}, 37,  145,
  2000.
\newblock http://planck.esa.int.

\bibitem[Park et~al.(2007)Park, Park, and III]{Park:SILC400}
Chan-Gyung Park, Changbom Park, and J.~Richard~Gott III.
\newblock Cleaned three-year wmap cmb map: Magnitude of the quadrupole and
  alignment of large scale modes.
\newblock {\em Astrophys. J.}, 660, 959, 2007.
\end{thebibliography}
\end{document}